\documentclass[aps,pra,showpacs,twocolumn,amsmath,amssymb,superscriptaddress, footinbib]{revtex4}

\usepackage[english]{babel}

\usepackage{latexsym}
\usepackage{graphicx}
\usepackage{subfigure}
\usepackage{epsfig}
\usepackage{amsfonts}
\usepackage{amssymb}
\usepackage{amsmath}
\usepackage{bm, bbm}

\usepackage{lipsum} 
\usepackage[draft]{todonotes}   

\begin{document}

\title{Exploring the impact of fluctuation-induced criticality on non-hermitian skin effect and quantum sensors}

\author{Clement Ehrhardt}
\affiliation{Department of Physics,
Stockholm University, AlbaNova University Center, 106 91 Stockholm,
Sweden}

\author{Jonas Larson}
\affiliation{Department of Physics,
Stockholm University, AlbaNova University Center, 106 91 Stockholm,
Sweden}

\date{\today}

\begin{abstract}
In this paper, we present a concrete example that highlights how predictions in non-Hermitian quantum mechanics can be inaccurately influenced by the absence of environment-induced fluctuations in the model. Specifically, we investigate the non-Hermitian skin effect and sensor in the Hatano-Nelson model, contrasting it with a more precise Lindblad description. Our analysis reveals that these phenomena can undergo breakdown when environmental fluctuations come to the forefront, resulting in a non-equilibrium phase transition from a localized skin phase to a delocalized phase. Beyond this specific case study, we engage in a broader discussion regarding the interpretations and implications of non-Hermitian quantum mechanics. This examination serves to broaden our understanding of these phenomena and their potential consequences.
\end{abstract}

\maketitle

\section{Introduction} 
In recent years, the field of non-Hermitian (NH) quantum mechanics (QM) has experienced a remarkable resurgence~\cite{ueda}. This renaissance can be traced back to the intriguing discovery that $\mathcal{PT}$-symmetric Hamiltonians, not necessarily Hermitian, can yield real spectra~\cite{bender}. A pivotal moment in this revival occurred with the introduction of biorthogonal QM~\cite{bioqm}, which ignited debates about the fundamental nature of QM. It challenged the long-held notion that observables must be represented solely by Hermitian operators~\cite{largeqm}.

The focus of NH QM has evolved to explore novel phenomena that emerge when we relax the constraints of Hermiticity and unitarity. One of the most extensively studied phenomena is the NH skin effect~\cite{skinref,skineffect,skineffect2}, which renders extreme sensitivity to non-local perturbations~\cite{ee,expsens,sensor}. For certain NH local Hamiltonians with open boundary conditions, all left/right eigenvectors $|\phi_n^{L,R}\rangle$ localize to one of the edges, offering intriguing possibilities for detection of weak signals~\cite{sensor,sensor4}.

NH QM often serves as an effective description of open quantum systems, typically arising from the interaction with an external environment. However, this approach raises questions about the treatment of fluctuations and the potential violation of well-established quantum theorems~\cite{nosignaling,nocloning,LR,entincrease,unred}.

This paper adopts a different perspective by employing the Lindblad master equation (LME) as a foundational framework to analyze quantum systems exposed to losses. Unlike NH QM, we do not neglect fluctuations, thus avoiding concerns related to quantum jumps. We also explore the implications and interpretations of NH theories in greater detail. Our study focuses on a specific example, where fluctuations qualitatively alter the physics of the system. We investigate a LME that reduces to the Hatano-Nelson (HN) model~\cite{hn} in the absence of quantum jumps, revealing a breakdown of the NH skin effect in favor of a delocalized phase. We discuss how such non-equilibrium criticality relates to earlier models in the context of optical bistability~\cite{optbis,optbis2}.

In conclusion, we examine the effects of fluctuations on NH QM, offering a perspective that complements existing research~\cite{comp1,physstate,nhlindblad1,nhlindblad2,nhlindblad2b,hatanonelsonlindblad,nhlindblad4,sensor2,jumpperturb}, especially by identifying a phenomenon of fluctuation-induced criticality which qualitatively alters the physical properties. We aim to provide a more detailed understanding of the role of fluctuations in NH QM. This will help shed light on the applicabilities of the theory in the quantum regime.

The paper is structured as follows: In the next section, we provide an in-depth discussion of non-unitary time evolution, with a particular focus on its description within the LME. We emphasize the importance of CPTP (Completely Positive, Trace-Preserving) maps and use them to argue why eigenvectors of a Liouvillian should not be considered as physical states. In Sec.~\ref{sec3}, we introduce the model system, the HN model in Subsec.~\ref{sec3a}, and its LME realization in Subsec.~\ref{hnreal}. Our main findings are presented in Sec.~\ref{sec4}, beginning with an exploration of the NH skin effect in  Subsec.~\ref{skineffectsec} and then a discussion of how this translates to applications in sensing in Subsec.~\ref{sensorsec}. We conclude with a discussion in Sec.~\ref{sec:con}. Additionally, we include two appendices. The first provides general comments on open quantum systems (appendix.~\ref{sec:app1}), and in the second, we demonstrate how our results for the HN model also apply to the NH SSH model (appendix.~\ref{sec:app2}).

\section{Non-unitary quantum evolution}
In this section, our primary aim is not to present new findings but rather to provide context by discussing general aspects of quantum state evolution. We specifically focus on the evolution generated by the LME. It is important to note that the Liouvillian, which is responsible for governing time evolution within the LME framework, is not represented by an observable. This distinction leads to significant differences compared to Hamiltonian systems. For instance, the eigenvectors of the Liouvillian do not typically represent physical states.

Nonetheless, a NH Hamiltonian somewhat falls between a traditional Hamiltonian and a master equation, but the field's terminology tends to lean more towards that of a Hamiltonian system. Having addressed these formal issues, the subsequent section will explore a concrete example as we apply our knowledge to the NH mode.

\subsection{The Lindblad master equation}
In Appendix~\ref{sec:app1}, we provide a more detailed description of open quantum systems. In this section, we will simply state that our system, denoted as $\mathcal{S}$, is weakly coupled to its surrounding environment. This inevitably implies that the evolution of the system alone cannot be described solely through unitary time evolution. However, we can still assume that the state $\hat\rho(t)$, which characterizes the physical properties of the system, adheres to the following physical state conditions:
\begin{equation}\label{phys}
\begin{array}{lllll}
(i) & \quad &\mathrm{Tr}\left[\hat\rho(t)\right]=1, &\qquad & \mathrm{Normalization},\\ \\ 
(ii) & \quad &\hat\rho(t)=\hat\rho^\dagger(t), &\qquad & \mathrm{Hermiticity},\\  \\
(iii) & \quad &|\!|\hat\rho(t)|\!|\geq0, &\qquad & \mathrm{Positivity}.
\end{array}
\end{equation}
The first condition corresponds to standard normalization, which preserves probabilities. The second condition ensures that all eigenvalues of $\hat\rho(t)$ are real, while the third condition guarantees that all eigenvalues are non-negative. These conditions are crucial for maintaining the probability interpretation of quantum mechanics, e.g. avoidning negative probabilities. It is important to note that these conditions must hold for all times, $t$. A mapping of physical states into new physical states, hence obeying the above conditions, is referred to as a completely positive trace-preserving map (CPTP)~\cite{nc,preskill}. In classical systems that emulate quantum dynamics (for example after applying the paraxial approximation to light propagation in non-linear media), deviations from the first condition are analogous to the loss or gain of particles or intensity, leading to a departure from the probabilistic nature.

Lindblad posed the question~\cite{ingemar}: What is the most general differential equation describing the time evolution of a quantum state, ensuring that the evolved state $\hat\rho(t)$ remains a valid density operator at all times? The most general form of such a dynamical CPTP map (on a differential structure) can be expressed in the Lindblad form~\cite{lindblad}
\begin{equation}\label{lind}
\begin{array}{lll}
\displaystyle{\frac{\partial}{\partial t}}\hat\rho & = & \hat{\mathcal{L}}\left[\hat\rho\right]
= \displaystyle{i\left[\hat\rho,\hat H\right]+\hat{\mathcal{D}}\left[\hat\rho\right]}\\ \\
 &  = & \displaystyle{i\left[\hat\rho,\hat H\right]+\sum_k\gamma_k\left(2\hat L_k\hat\rho\hat L_k^\dagger-\hat L_k^\dagger\hat L_k\hat\rho-\hat\rho\hat L_k^\dagger\hat L_k\right)}.
\end{array}
\end{equation}
Here, we introduce the Liouvillian operator $\hat{\mathcal{L}}$, and the dissipator operator $\hat{\mathcal{D}}$ accounts for the influence of the environment. The $\gamma_k$ values represent the ``decay rates'' for channel $k$, and the $\hat L_k$'s are the Lindblad jump operators~\cite{bp}. The above LME can be put on the form
\begin{equation}\label{lind2}
\displaystyle{\frac{\partial}{\partial t}}\hat\rho=\hat{\mathcal{L}}_c\left[\hat\rho\right]\equiv i\left(\hat\rho\hat H_\mathrm{eff}-\hat H_\mathrm{eff}^\dagger\hat\rho\right)+\hat{\mathcal{J}}_c\left[\hat\rho\right],
\end{equation}
with the effective NH ``Hamiltonian'' defined as
\begin{equation}\label{nhh}
\hat H_\mathrm{eff}=\hat H-i\sum_k\gamma_n\hat L_k^\dagger\hat L_k,
\end{equation}
and the jump super-operator
\begin{equation}
\hat{\mathcal{J}}_c\left[\hat\rho\right]=2c\sum_n\gamma_n\hat L_n\hat\rho\hat L_n^\dagger.
\end{equation}
Please note that we use ``Hamiltonian'' in quotation marks because, in general, it does not correspond to a traditional physical Hamiltonian. Specifically, when using ``Hamiltonian'' we refer to Eq.~(\ref{nhh}). The subscript $0\leq c\leq1$ parametrizes the master equation, where $c=1$ reproduces the correct LME~(\ref{lind}), and for $c=0$, the evolution is governed by the NH ``Hamiltonian'' $\hat H_\mathrm{eff}$. We will exclusively focus on time-independent jump operators. It is worth mentioning that a microscopic derivation of the LME~(\ref{lind}) typically relies on three approximations~\cite{bp,carmichael}: the Markovian, Born, and secular approximations. The properties of the environment and the system-environment Hamiltonian determine the dissipator $\hat{\mathcal{D}}\left[\hat\rho\right]$, including the rates $\gamma_k$ and the jump operators $\hat L_k$. Importantly, in especially cold atom and optical systems, it is feasible, to a high degree, to engineer both the system and its coupling to the environment to achieve desired Liouvillians~\cite{lengineer}. In the following section, we provide explicit suggestions on how to utilize relaxation to implement a HN-like model characterized by unbalanced left/right hopping in a 1D tight-binding lattice.
 
\subsection{Some general properties of the Lindblad master equation}
An eigenvector, denoted as $\hat\rho_j$ and its corresponding eigenvalue, $\mu_j$, of the LME are defined by the equation~\cite{unik,albert,patrik}
\begin{equation}\label{leigs}
\hat{\mathcal{L}}\left[\hat\rho_j\right]=\mu_j\hat\rho_j.
\end{equation}
In principle, we should refer to $\hat\rho_j$ as an ``eigenmatrix'', but for simplicity, we will continue to use the term ``eigenvector''. This choice is justified since the LME~(\ref{lind}) can be vectorized as
\begin{equation}\label{lindbladvec}
    \frac{d}{dt}|\rho\rangle\!\rangle=\hat{\mathcal{L}}_v|\rho\rangle\!\rangle
\end{equation}
with the voctorized Liouvillian, $\hat{\mathcal{L}}_v$, defined as
\begin{equation}
\begin{array}{lll}
    \displaystyle{\hat{\mathcal{L}}_v} & = & -i\left(\hat H\otimes\mathbb{I}+\mathbb{I}\otimes\hat H^T\right)\\ \\
    & & +\displaystyle{\sum_k\gamma_k\!\left[2\hat L_k\!\otimes\hat L_k^{\dagger T}-\hat L_k^\dagger\hat L_k\!\otimes\mathbb{I}-\mathbb{I}\otimes\!\left(\hat L_k^\dagger\hat L_k\right)^T\right]\!.}
    \end{array}
\end{equation}
Now, given a finite Hilbert space dimension $D(\mathcal{H})=N$, $\hat\rho$ is represented as an  $N^2$-component vector, $|\rho\rangle\!\rangle$, rather than an $N\times N$ matrix~\cite{vec}.  The vector  $|\rho\rangle\!\rangle$ resides in a Liouville space $\mathcal{L}$, which is the direct product of two Hilbert spaces, such that the dimension of the Liouville space is $D(\mathcal{L})=N^2$.

For the numerical computations presented in Sec.~\ref{sec4}, we utilize the vectorized version of the LME. We select an appropriate (Fock) basis and express the operators  $\hat H$ and $\hat L_k$ as matrices, which are employed to construct the Liouvillian matrix $\hat{\mathcal{L}}_v$. Subsequently, we numerically diagonalize it to determine the Liouvillian spectrum $\mu_j$ and corresponding eigenvectors $|\rho_j\rangle\!\rangle\leftrightarrow\hat\rho_j$ (including the steady state). The steady state of Eq.~(\ref{leigs}) complies with $\hat{\mathcal{L}}\left[\hat\rho_\mathrm{ss}\right]=0$. It is important to note that the vector components of  $|\rho\rangle\!\rangle$ do not represent probability amplitudes. For the scalar product, we have  $\langle\!\langle\rho|\varrho\rangle\!\rangle=\mathrm{Tr}\left[\hat\rho\hat\varrho\right]$.

The uniqueness of the steady state $\hat\rho_\mathrm{ss}$ has been extensively discussed in previous works, starting with studies by Spohn and continued by others; refer, for example, to Refs.~\cite{unik}. For LMEs possessing symmetries, multiple steady states can arise, leading to non-trivial situations reminiscent of spontaneous symmetry breaking, akin to continuous phase transitions in closed systems~\cite{albert,patrik}. We will return to this in the next section when analyzing a LME-extension of the HN model. 

The LME is a CPTP map, i.e. $dN/dt=0$ where $N=\mathrm{Tr}\left[\hat\rho(t)\right]$, and $\langle\psi|\hat\rho(t)|\psi\rangle\geq0$ for any state $|\psi\rangle$.  However, it is important to note that as soon as the previously introduced parameter $c\neq1$, the CPTP property is generally lost. Specifically, in the NH limit with $c=0$, it is only under very special circumstances that the evolution remains trace-preserving, even when the spectrum is purely real. A significant consequence of the CPTP property of the LME can be directly inferred~\cite{physstate,sofia}. First, we observe that an eigenvector evolves as 
\begin{equation}
    \hat\rho_j(t)=e^{\mu_jt}\hat\rho_j,
\end{equation}
and thus, its trace 
\begin{equation}
    N_j(t)=\mathrm{Tr}\left[\hat\rho_j(t)\right]=e^{\mu_jt}\mathrm{Tr}\left[\hat\rho_j\right].
\end{equation}
However, since $dN(t)/dt=0$, we must have  $\mathrm{Tr}\left[\hat\rho_j\right]=0$ whenever $\mu_j\neq0$. In other words, every eigenvector, apart from the steady states, is traceless. According to~(\ref{phys}), this leads to the crucial result: 
\begin{enumerate}
  \item[] {\it Every eigenvector $\hat\rho_j$ of some Lindblad Liouvillian $\hat{\mathcal{L}}$, apart for its steady state(s), are unphysical.}
\end{enumerate}

For classical systems, it is well-established that only the steady state of a master equation can exclusively consist of non-negative entries. In contrast, all other eigenvectors possess at least one entry that is negative, rendering them unsuitable for representing physical states, as these entries correspond to probabilities~\cite{vk}. The result stated above parallels this observation, but now applies to quantum systems. Consequently, this motivates us to express the following:
\begin{enumerate}
  \item[] {\it As with a master equation, the eigenvectors and eigenvalues of $\hat{\mathcal{L}}_c$ should not, in the general case, be interpreted as representing physical states and observable quantities, respectively.}
\end{enumerate}
This assertion is not confined solely to the LME case ($c=1$), as we contend (as elaborated below) that the evolution generated by $\hat H_\mathrm{eff}$ should not be regarded as Hamiltonian time evolution. When one unravels the Lindblad master equation in quantum trajectories~\cite{unraveling}, the NH (renormalized) time evolution arises through post-selection. Thus, in this case, while it may seem as if $\hat H_\mathrm{eff}$ generates the time evolution, one should not forget that it is actually only true due to the post-selection constraint. 

In both classical and quantum scenarios, the eigenvectors still constitute a complete set, apart at exceptional points, meaning that any other state can be expressed as a linear combination of them, $\hat\rho=\sum_jp_j\hat\rho_j$, using coefficients $p_j$. Thus, the time-evolved state can be expressed as
\begin{equation}\label{linsum}
\hat\rho(t)=\hat\rho_\mathrm{ss}+\sum_jp_je^{\mu_jt}\hat\rho_j,
\end{equation}
where it is implied that the sum excludes eigenvectors with $\mu_j=0$. To ensure CPTP behavior, the eigenvalues must satisfy $\mathrm{Re}(\mu_j)\leq0$~\cite{bp,huelga,nonneg}. Consequently, all states $\hat\rho_j$ with a non-zero real part $\mathrm{Re}(\mu_j)$ are exponentially suppressed as time progresses. The Liouvillian gap, defined as~\cite{lgap}
\begin{equation}\label{lgap}
    \Delta_\mathcal{L}=\min_j\mathrm{Re}(\mu_j),
\end{equation}
can be regarded as providing an initial estimate for the relaxation time scale towards the steady state (although the scenario can be more intricate~\cite{lgap2}). 

In the context of quantum phase transitions, in the thermodynamic limit, the ground state exhibits non-analytic behavior at the critical point, and additionally, the spectrum necessarily becomes gapless at this juncture. Analogously, the steady state $\hat\rho_\mathrm{ss}$ may display similar non-analytic behavior, accompanied by the vanishing of the Liouvillian gap~\cite{lgap,lcrit}. It is important to note that the steady state does not decay, although under certain conditions~\cite{nons}, $\mathrm{Re}(\mu_j)$ may equal 0 while $\mathrm{Im}(\mu_j)\neq 0$, potentially resulting in non-stationary states. However, this requires that the eigenvalues $\mu_j$ must appear in complex conjugate pairs if they possess a non-vanishing imaginary part.

Returning to Eq.~(\ref{lind2}), if we put the last term to zero (i.e. $c=0$) the evolution is governed by the NH ``Hamiltonian'' $\hat H_\mathrm{eff}$ of Eq.~(\ref{nhh}). In this case, let us assume that we know the right eigenvectors 
\begin{equation}
    \hat H_\mathrm{eff}^\dagger|\varphi_l^R\rangle=\nu_n|\varphi_l^R\rangle.
\end{equation}
That is, the left eigenvectors obey 
\begin{equation}
    \hat H_\mathrm{eff}|\varphi_l^L\rangle=\nu_n^*|\varphi_l^L\rangle.
\end{equation}
The eigenvalues and eigenvectors of $\hat{\mathcal{L}}_{c=0}$ become 
\begin{equation}\label{nheigs}
    \mu_j^0=i(\nu_l^*-\nu_k),\hspace{1cm}\hat\rho_j^0=|\varphi_k^R\rangle\langle\varphi_l^R|.
\end{equation}
Here, the superscript 0 denotes the case $c=0$, and $j$ replaces the double indices $(l,k)$. As mentioned earlier, if the Hilbert space dimension is finite, $D(\mathcal{H})=N$, the Liouville space dimension is $N^2$, as seen in Eq.~(\ref{nheigs}) since $1\leq k,\,l\leq N$ implies $1\leq j\leq N^2$. Thus, we have far more eigenvectors/values of the Liouvillian $\hat{\mathcal{L}}_{c=0}$ than for the NH ``Hamiltonian'' $\hat H_\mathrm{eff}$. 

We can parametrize the eigenvectors of $\hat{\mathcal{L}}_{c}$ with the number $c$ (i.e. $\hat\rho_j^c$), such that $\hat\rho_j^{c=0}$ reproduces the eigenvectors in (\ref{nheigs}), while $\hat\rho_j^{c=1}$ provides those of Eq.~(\ref{leigs}). For finite dimensionsional Hilbert spaces we expect the vectors $\hat\rho_j^c$ and eigenvalues $\mu_j^c$ to be analytic in $c$, with the exceptions at possible exceptional points~\cite{ep}. Letting $c = 0$, if $\nu_l$ is real, it follows that the vectors $\hat\rho_j^0=|\varphi_l^R\rangle\langle\varphi_l^R|$ are steady states and also physical. However, if  $\mathrm{Im}(\nu_j)\neq0$, these eigenvectors $\hat\rho_j^0=|\varphi_l^R\rangle\langle\varphi_l^R|$ are no longer steady states since their norms are not preserved.

Let us provide some context for the discussion above. As mentioned in the introduction, NH QM challenges well-established physical concepts. Furthermore, an ongoing debate surrounds the clear interpretation of the theory emerging from NH QM, particularly with regard to which states should be employed to describe time evolution~\cite{hatanonelsonlindblad,rrdebate1}.

One approach to circumvent potential issues, such as violating the no-signaling theorem, is the introduction of biorthogonal QM~\cite{nogo,bioqm}. This approach hinges on the biorthogonality property, which allows for the construction of mutually orthogonal left $|\varphi_j^L\rangle$ and right $|\varphi_j^R\rangle$ eigenvectors, such that $\langle\varphi_l^L|\varphi_j^R\rangle=\delta_{lj}$. For example, this property leads to the modified resolution of identity, which becomes $\hat O=\sum_jo_j|\phi_j^R\rangle\langle\phi_j^L|$. Additionally, the spectral resolution of an operator can be expressed as $\hat O=\sum_jo_j|\phi_j^R\rangle\langle\phi_j^L|$, where $o_j$ and $|\phi_j^R\rangle$ ($\langle\phi_j^L|$) represent the eigenvalue and right (left) eigenvectors of the operator, respectively. In the biorthogonal framework, assuming an initial state $|\psi(0)\rangle$, the state at a later time is described by two vectors: $|\psi^{L,R}(t)\rangle$, referred to as the `left' and the `right' evolved state. Expectations of observables $\hat O$ should also be evaluated according to this `state,' given by 
\begin{equation}
\mathcal{O}_{|\psi^{L,R}\rangle}(t)=\langle\psi^L(t)|\hat O|\psi^R(t)\rangle=\mathrm{Tr}\left[\hat O\hat\rho(t)\right],
\end{equation}
with $\hat\rho(t)=|\psi^R(t)\rangle\langle\psi^L(t)|$.

This can be extended further; the expectation of an operator for the $j$'th eigenvector becomes $\mathcal{O}_j=\langle\varphi_j^L|\hat O|\varphi_j^R\rangle$. When applied to the position operator $\hat n_n=|n\rangle\langle n|$ (where $|n\rangle$ represents the particle localized to site $n$) of the HN model, one finds that the ``biorthogonal'' eigenvectors are not localized at the edges but instead are delocalized within the bulk~\cite{el2} (see Eq.~(\ref{eigvec1}) below for the eigenvectors).

Of course, if the generator of time evolution is a Hermitian Hamiltonian, this reproduces standard quantum mechanics since $\langle\psi^L|$ will equal the right bra vector $\langle\psi^R|$. However, while the biorthogonal approach seems to resolve some issues, it comes with caveats. First, we note that $\langle\varphi_l^L|\varphi_j^R\rangle=\delta_{lj}$ does not set the norm of the left/right eigenvectors independently of each other. In fact, this results in a family of scalar products parametrized by a metric $\eta$~\cite{nogo,elisabet}. Thus, an ambiguity, similar to a gauge freedom, arises. It has been argued that for a given Hamiltonian, there is a preferable metric to be used, which, however, implies that the metric, and thereby the scalar product, changes when, for example, you add one particle or site (modifying the Hilbert space dimension) to your system~\cite{bioqm,elisabet}.

Secondly, the fact that a physical state must be ascribed both a bra- and ket-vector is non-intuitive using common knowledge, and we have not even addressed mixed states. Now, it has been argued that the biorthogonal scalar product is not the one to be used in order to describe the time evolution of actual physical systems~\cite{rrdebate1,hatanonelsonlindblad}. This aligns with our description. If we take the LME as our starting point and think that we can, at least in theory, connect it to NH QM by `turning off' the jump terms, we should not end up with the biorthogonal left/right formalism, but rather with a right/right (or left/left) density operator~\cite{com1}. In the post-selection approach, it is yet again the right/right state (complemented with renormalization) that describes the system.

It turns out that the spectral properties~(\ref{nheigs}) of the NH Hamiltonian  $\hat H_\mathrm{eff}$ can capture the full Liouvillian spectrum. To be more precise, under specific conditions, the spectrum of $\hat{\mathcal{L}}_{c=0}$ can be directly mapped to the spectrum of $\hat{\mathcal{L}}_{c=1}$~\cite{lspec}. This mapping becomes feasible when the system Hamiltonian supports particle conservation, as indicated by $\left[\hat N,\hat H\right]=0$, and every jump operator adheres $\left[\hat L_k,\hat N\right]=\hat L_k$. This situation is relevant in the context of spontaneous decay~\cite{bp}. It is worth noting that our Lindblad representation of the HN model, as presented in the following section, does not satisfy this second condition, and therefore, we cannot utilize such a property. 

Another interesting scenario arises when the Liouvillian has a quadratic dependence on the creation/annihilation operators $\hat a_k^\dagger/\hat a_k$~\cite{prosen}. Similar to quadratic Hamiltonians, the quadratic form of the Liouvillian can be employed for diagonalization through a generalized Bogoliubov–de Gennes approach. This method has been recently applied in the investigation of the skin effect in Liouvillians~\cite{nhlindblad2b}. However, it is important to note that the physical realization of the HN model we have in mind, as described in Subsec.~\ref{hnreal}, does not fall within this class of solvable models.

\section{Model system: Liouvillian for the Hatano-Nelson model}\label{sec3}
\subsection{The Hatano-Nelson model}\label{sec3a}
A fundamental model often used in the study of NH QM is the one proposed by Hatano and Nelson. This model describes a single particle within a one-dimensional tight-binding lattice of N sites, where the hopping between sites is unbalanced in the left and right directions~\cite{hn,hn2}. For open boundary conditions (BC), the model is represented by the (NH) Hamiltonian as
\begin{equation}\label{hnham}
    \hat H_\mathrm{HN}=\sum_{n=1}^{N-1}\left[\left(1-\delta\right)\hat a_n^\dagger\hat a_{n+1}+(1+\delta)\hat a_{n+1}^\dagger\hat a_n\right].
\end{equation}
Here, $\hat a_n$ ($\hat a_n^\dagger$) represent the annihilation (creation) operators for a particle at site $n$, satisfying the single particle constraint $\hat N=\sum_{n=1}^N\hat a_n^\dagger\hat a_n\equiv\sum_{n=1}^N\hat n_n=1$. The parameter $\delta$ varies between 0 and 1 and indicates the degree of asymmetry in the hopping to the left and right.

For a single particle, the model can also be expressed using bracket notation, for example, $\hat a_{n+1}^\dagger\hat a_n\leftrightarrow|n+1\rangle\langle n|$, and so forth, where $|n\rangle$ represents the number state of the particle localized at the $n$-th site. The model is typically presented in dimensionless units, scaled with respect to the symmetric hopping amplitude. It can be convenient to decompose the Hamiltonian into `real' and `imaginary' parts as follows
\begin{equation}\label{realham}
    \hat H_\mathrm{HN}=\hat H_R+i\delta\hat H_I,
\end{equation}
where 
\begin{equation}
    \hat H_R=\sum_{n=1}^{N-1}\left(\hat a_n^\dagger\hat a_{n+1}+\hat a_{n+1}^\dagger\hat a_n\right)
\end{equation}
and 
\begin{equation}
    \hat H_I=i\sum_{n=1}^{N-1}\left(\hat a_n^\dagger\hat a_{n+1}-\hat a_{n+1}^\dagger\hat a_n\right)
\end{equation}
Both $\hat H_R$ and $\hat H_I$ are hermitian. Notably, in the case of open boundary conditions, the commutator $\left[\hat H_R,\hat H_I\right]$ vanishes, except for the first and last diagonal elements.

For future reference, it is informative to introduce the ladder operators, denoted as $\hat E=\sum_n\hat a_n^\dagger\hat a_{n+1}$ and its hermitian conjugate $\hat E^\dagger$. These operators satisfy the following relationships: $\hat E|n\rangle=|n-1\rangle$ and $\hat E^\dagger|n\rangle=|n+1\rangle$ (it should be noted that for a finite lattice, $\hat E|1\rangle=0$ and $\hat E^\dagger|N\rangle=0$). When combined with the operator $\hat E_0$, which follows the relation $\hat E_0|n\rangle=n|n\rangle$, these three operators collectively constitute what is known as the Euclidean algebra~\cite{euclid}
\begin{equation}\label{ealg}
    \left[\hat E,\hat E_0\right]=-\hat E,\hspace{0.5cm}\left[\hat E^\dagger,\hat E_0\right]=\hat E^\dagger,\hspace{0.5cm}\left[\hat E,\hat E^\dagger\right]=0.
\end{equation}
Alternatively, we can introduce the ``psoition'' and ``momentum'' operators $\hat E_x=\hat E+\hat E^\dagger$ and $\hat E_p=i\left(\hat E-\hat E^\dagger\right)$, and the ``Hamiltonian'' reads $\hat H_\mathrm{HN}=\hat E_x+i\delta\hat E_p$.

For periodic BC, the real and imaginary parts commute and the eigenvectors are delocalized bulk states. For open BC the spectrum reads~\cite{ee,hnspec}
\begin{equation}\label{obc}
    \nu_j=2\sqrt{1-\delta^2}\cos(j\pi/(N+1)),
\end{equation}
while for periodic BC one has
\begin{equation}\label{pbc}
    \nu_j=2\left[\cos(2\pi j/N)-i\delta\sin(2\pi j/N)\right],
\end{equation}
where $j=1,2,\dots,N$. Consequently, the spectrum is purely real for open boundary conditions. However, the right and left eigenvectors are not orthogonal, which leads to non-unitary time evolution. All right eigenvectors, while not normalized, are exponentially localized towards the right edge and can be represented in terms of the number states as:
\begin{equation}\label{eigvec1}
    |\varphi_j^R\rangle=\sum_{n=1}^N\left(\frac{1+\delta}{1-\delta}\right)^n\sin\left(\frac{nj\pi}{N+1}\right)|n\rangle.
\end{equation}
The left eigenvectors can be obtained by substituting $\delta$ with $-\delta$. By using this substitution, we find that the biorthogonal QM provides the eigenvector site occupations $P_j(n)=\langle\varphi_j^L|\hat n_n|\varphi_j^R\rangle\propto\sin^2\left(\frac{nj\pi}{N+1}\right)$. The bulk nature of these states, when combining left and right eigenvectors, was previously observed by Hatano and Nelson~\cite{hn2}.

\subsection{Lindblad implementation of the Hatano-Nelson model}\label{hnreal}
Our objective is to construct jump operators $\hat L_k$, which enable the NH ``Hamiltonian'' in Eq.~(\ref{nhh}) to match the HN Hamiltonian~(\ref{hnham}). Once these operators are identified, we propose a physical system where this can be implemented. By separating the real and imaginary components of both equations, we should obtain the following relationships
\begin{equation}
    \hat H_R=\hat H
\end{equation}
and
\begin{equation}\label{decomp}
    \delta\hat H_I=-\sum_k\gamma_k\hat L_k^\dagger\hat L_k.
\end{equation}
The first equation implies that our Hamiltonian should resemble a simple $N$-site tight-binding chain. The second identity is more complex, as the spectrum of  $\hat H_I$ is not strictly non-negative, while $\hat L_k^\dagger\hat L_k$  is positive semi-definite (recall $\gamma_k\geq0$). Nonetheless, we can resolve this issue by shifting the entire spectrum of the NH ``Hamiltonian'' by an imaginary constant of  $2i\delta$~\cite{hatanonelsonlindblad}. It is worth noting that the decomposition into jump operators is not unique; instead, there are numerous possible combinations. To illustrate this, the right-hand side of equation~(\ref{decomp}) can be compactly expressed as $-\hat{\bf L}^\dagger\Gamma\hat{\bf L}$, where $\hat{\bf L}=(\hat L_1,\,L_2,\dots)^t$ is a column vector containing all jump operators (for a finite dimension $D(\mathcal{H})=N$, the number of independent jump operators can be limited to $N^2-1$), and $\Gamma$ is a diagonal matrix with $\gamma_k$ along its diagonal. It is evident that this term remains invariant under some unitary transformation $\hat U$, i.e., $\hat{\bf L}'=\hat U\hat{\bf L}$ and $\Gamma'=\hat U\Gamma\hat U^{-1}$. The specific choice should be determined by the physical system in question. Two cases, in particular, come to mind: 

\begin{enumerate}
  \item {\bf Local decay channels.} At each site $n$, we can associate a jump operator $\hat L_n$. When this operator acts on the state $|n\rangle$, there is a non-negligible probability for the particle to transition to $|n+1\rangle$. A straightforward approach might be to set $\hat L_n=\hat a_n\hat a_{n+1}^\dagger$. However, this does not suffice, as $\hat L_n^\dagger\hat L_n$ results in $\hat n_n$, which cannot be used to construct the desired effective NH ``Hamiltonian''. Instead, we should employ the following expression:
  \begin{equation}\label{locjump}
      \hat L_n=i\hat a_n\hat a_{n+1}^\dagger+\hat n_n,
  \end{equation}
ensuring that the application of the jump operator to $|n\rangle$ leaves the state in a superposition of $|n\rangle$ and $|n+1\rangle$.

  \item {\bf Collective decay channels.} If the jump operator does not ``keep track'' of the particle's position, we say that the jump occurs independently of the particle's position. In this case, we sum the local jump operators into a single one, i.e.
  \begin{equation}\label{coljump}
      \hat L_c=i\sum_{n=1}^{N-1}\hat a_n\hat a_{n+1}^\dagger+\sum_{n=1}^N\hat n_n\equiv\hat E+\mathbb{I},
  \end{equation}
  where $\hat E$ was defined above Eq.~(\ref{ealg}).
\end{enumerate}
The two scenarios described above result in different physical realizations. Unraveling the LME in terms of stochastic 'quantum trajectories'~\cite{unraveling} provides a physical picture of the system's evolution. In summary, when we possess complete knowledge about the environment, the system should evolve deterministically with the NH ``Hamiltonian''~(\ref{nhh}). However, this evolution is occasionally interrupted by quantum jumps, whose effects are described by the application of jump operators to the state. This process involves the instantaneous renormalization of the state under non-unitary time evolution.

In practice, `keeping track' of the environment involves monitoring any photons spontaneously emitted by the system. If a photon is detected, we can infer that the system has undergone a stochastic jump. This provides a distinct physical differentiation between the two scenarios. In the latter situation, the emitted photon does not yield information about the particle's position. In contrast, in the former scenario, a recorded photon not only signifies that a jump has occurred but also reveals the particle's location, effectively collapsing the wave function within the lattice.

As a result, it becomes evident that we can anticipate different behaviors under the evolution driven by the different Lindblad operators. For the remainder of our discussion, we will focus on the collective decay channel. However, for a finite lattice comprising $N$ sites, we should note that the jump operator $\hat L_c$ of Eq.~(\ref{coljump}) is insufficient. We need to supplement it with an additional dephasing operator $\hat L_1=\hat a_1^\dagger\hat a_1$ for the first site. With this consideration in mind we have
\begin{equation}
    \hat L_c^\dagger\hat L_c+\hat L_1^\dagger\hat L_1=\hat H_I+2\mathbb{I}, 
\end{equation}
and by further identifying $\gamma_c=\gamma_1\equiv\gamma=-\delta$ we have our LME corresponding to the HN model
\begin{equation}\label{collind}
\begin{array}{lll}
    \displaystyle{\frac{\partial}{\partial t}}\hat\rho & = & \displaystyle{i\left[\hat\rho,\hat H\right]+\gamma\left(2\hat L_c\hat\rho\hat L_c^\dagger-\hat L_c^\dagger\hat L_c\hat\rho-\hat\rho\hat L_c^\dagger\hat L_c\right)}\\ \\
    & & +\displaystyle{\gamma\left(2\hat L_1\hat\rho\hat L_1^\dagger-\hat L_1^\dagger\hat L_1\hat\rho-\hat\rho\hat L_1^\dagger\hat L_1\right)},
 \end{array}
\end{equation}
where the Hamiltonian $\hat H$ identifies the real part~(\ref{realham}) of the HN Hamiltonian.

\begin{figure}[h]
\centerline{\includegraphics[width=6cm]{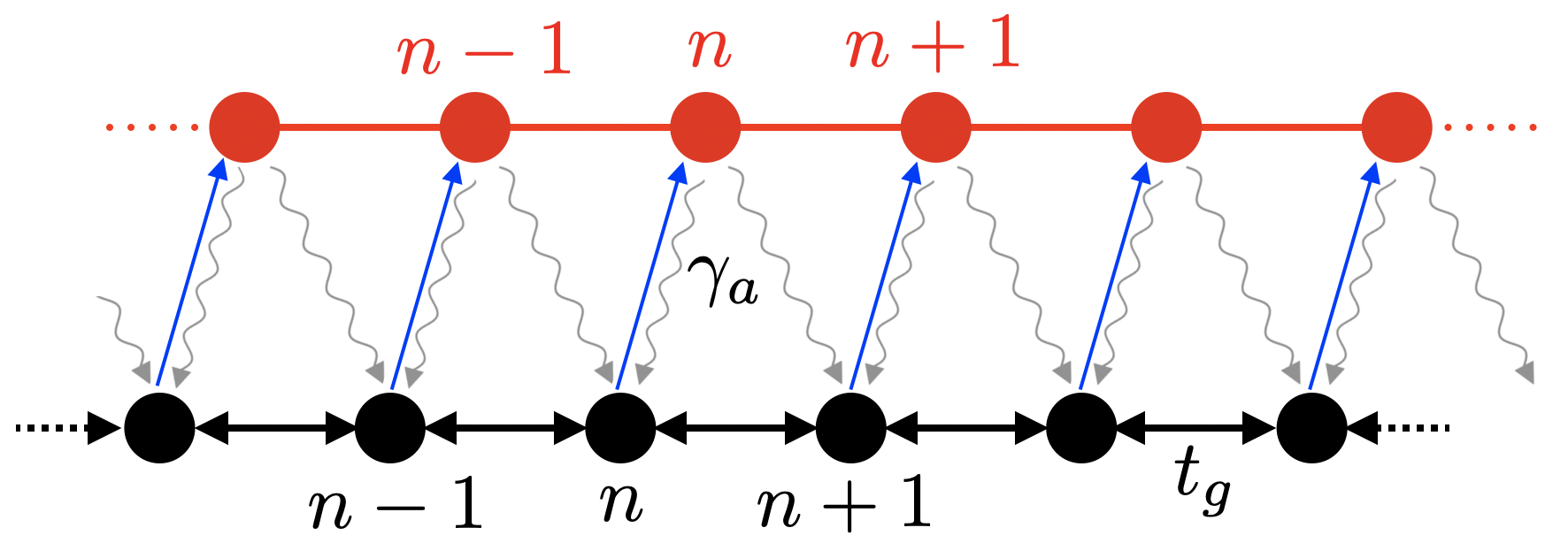}}
\caption{Illustration of the Implementation of the HN Model as an open quantum system. In this schematic representation, atomic lattice sites are depicted as black and red dots, signifying two distinct internal hyperfine levels $|g\rangle$ and $|e\rangle$. These atoms are confined within an optical lattice, which is state-dependent. The lattice is designed so that the two internal states experience slight shifts concerning each other. Additionally, the excited atomic state $|e\rangle$ is inherently unstable and quickly relaxes (with an atomic rate $\gamma_a$) back to the stable atomic state $|g\rangle$. As a result of this shift in the respective potential minima and the relaxation process, an effective drift occurs in the lattice, moving in the right direction. Specifically, when an atom in the excited state $|e\rangle$ at site $n$ , it relaxes back to the ground state $|g\rangle$ at either the site $n$ or $n+1$, while relaxation to other sites is negligible.} \label{fig1}
\end{figure}

In this section, we offer a conceptual overview of how this model could be implemented within a cold atom setup~\cite{coldatom}. A similar concept was recently proposed using motional sidebands in a trapped ion setup~\cite{trapsetup}. Our discussion centers on an atom that possesses two internal hyperfine levels: $|g\rangle$ (ground) and $|e\rangle$ (excited). This atom is confined within a one-dimensional lattice. Under the usual approximations, which include tight-binding and single-band assumptions, we implement a resonant classical drive between the two internal atomic states. This leads to the following second-quantized lattice Hamiltonian
\begin{equation}\label{fullham}
\begin{array}{lll}
    \hat H & = & \displaystyle{-t_g\sum_n\left(\hat a_n^\dagger\hat a_{n+1}+h.c.\right)-t_e\sum_n\left(\hat b_n^\dagger\hat b_{n+1}+h.c.\right)}\\ \\
    & & \displaystyle{+g\sum_n\left(\hat b_n^\dagger\hat a_{n}+h.c.\right)}.
    \end{array}
\end{equation}
Here, $t_{g}$ and $t_{e}$ represent the tunneling rates of the two atomic species, and $\hat a_n$ ($\hat a_n^\dagger$) and $\hat b_n$ ($\hat b_n^\dagger$) correspond to the single-site annihilation (creation) operators. The parameter $g$ signifies the effective Rabi coupling. Notably, the driving mechanism couples atomic internal states within a single site. This implies a smooth laser profile and a sufficiently deep lattice. For simplicity, we can assume $|t_e|\ll|t_g|$ to suppress lattice dynamics of the excited states. We normalize energies in terms of the tunneling rate $t_g$, setting $t_g=1$ from this point forward. If the $|e\rangle$-level rapidly relaxes to the ground state $|g\rangle$, adiabatic elimination is applicable~\cite{adel}. Consequently, we commence with the above Hamiltonian, complemented by a bath of oscillators inducing couplings between the two atomic levels. In this derivation, we employ the standard approximations, which include Born, Markov, and secular approximations~\cite{bp}. The physical setup is depicted in Figure 1. In the resulting LME for the atomic ground state $|g\rangle$, the effective ``decay rate'' $\gamma$ is directly proportional to the Rabi frequency $g$. It also significantly depends on the Franck-Condon factors, which are proportional to the overlaps between Wannier functions of the two species. The alignment of the two lattices and their lattice depths should ensure that a Wannier function localized to the $n$'th site of the ``red'' lattice primarily overlaps with Wannier functions at the  $n$'th and  $n+1$'th lattice sites of the ``black'' lattice. In this scenario, and under the assumption of precise phase alignment between the two decay channels (see Eq.~(\ref{coljump})), we achieve the desired LME.

\section{Case studies}\label{sec4}
In the section we consider two phenomena within the HN model: the skin effect~\cite{skinref,skineffect,skineffect2} and NH sensors~\cite{sensor,sensor2,sensor3}. The aim is to compare and contrast outcomes obtained using either the HN model of Eq.~(\ref{hnham}) or the full LME~(\ref{collind}). 

\begin{figure}[h]
\centerline{\includegraphics[width=8.4cm]{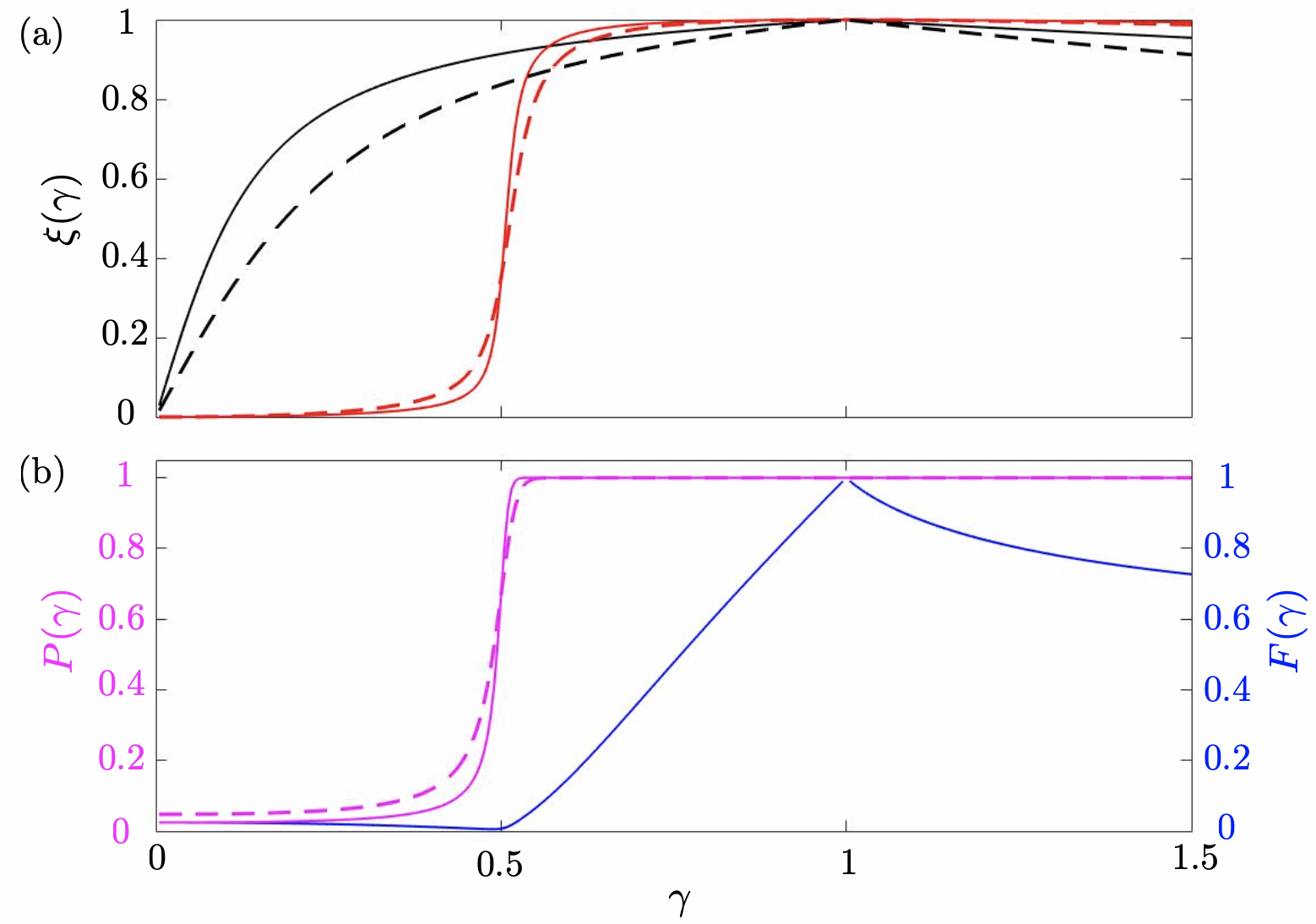}}
\caption{(Color online) In the upper plot (a) we display the scaled position~(\ref{pos}) for both the HN model (black), calculated as $\xi=\langle\varphi_N^R|\hat S|\varphi_N^R\rangle/N$, and the full LME (red). The solid lines represent a lattice of length $N=41$, while the dashed lines represent $N=21$. In the lower plot (b), we illustrate the purity~(\ref{pur}) of the steady state $\hat\rho_\mathrm{ss}$ (purple) and the fidelity~(\ref{fid}) between the HN ``ground state'' and the Lindblad steady states (blue). The key observation is the appearance of a critical point at $\gamma_c=1/2$ in the LME, which is absent in the HN model. Below the critical point, $\gamma<\gamma_c$, fluctuations become essential, leading to the destabilization of the skin effect and system delocalization. Above the critical point,$\gamma>\gamma_c$, the skin effect persists, even though the fidelity is not very high. } \label{fig2}
\end{figure}

\subsection{Skin effect}\label{skineffectsec}
While the phenomenon was initially discovered by Hatano and Nelson in the 1990s~\cite{hn,hn2}, the term ``non-hermitian skin effect'' was coined by Yau and Wang more than two decades later when they conducted a more in-depth and comprehensive study~\cite{skinref}. When considering a NH lattice ``Hamiltonian'' with open boundary conditions, the skin effect is characterized by the eigenvectors becoming localized at the edges. This has a profound consequence: the system becomes exponentially sensitive to non-local perturbations~\cite{hn2}. This sensitivity can be demonstrated by extending standard perturbation theory to NH models~\cite{sensor}. Hence, NH edge localization and exponential sensitivity are two sides of the same coin. We will revisit the latter concept in Subsec.~\ref{sensorsec}.

To characterize edge localization, we consider a lattice with $N=2n-1$ (where $n$ is a positive integer) sites and introduce the matrix $\hat S=\mathrm{diag}(-n:n)$ along with the scaled position expectation $\xi$ defined as
\begin{equation}\label{pos}
    \xi=\frac{1}{N}\mathrm{Tr}\left[\hat\rho_\mathrm{ss}\hat S\right].
\end{equation}
If $\xi=\pm1$, the state $\hat\rho$ is maximally localized at one of the edges, whereas $\xi=0$ indicates that the state is centered within the lattice. The uncertainty $\Delta\xi=\sqrt{\langle\hat S^2\rangle-\langle\hat S\rangle^2}$ determines the degree of localization of the state. Therefore, when $\Delta\xi$ is of the order of $N$, the state becomes delocalized over the entire lattice. The purity of the state is quantified by
\begin{equation}\label{pur}
    P=\mathrm{Tr}\left[\hat\rho_\mathrm{ss}^2\right].
\end{equation}
Notably, in the context of NH QM, the (normalized) eigenvectors are pure, and their normalization ensures that $P_\mathrm{NH}=1$.

In the thermodynamic limit of an infinite lattice, where $\left[\hat L_c,\hat L_c^\dagger\right]=0$, the steady state of the LME simplifies to the maximally mixed state, $\hat\rho_\mathrm{ss}\propto\mathbb{I}$~\cite{patrik}. However, for a finite lattice, non-trivial steady states can emerge, particularly those localized at the edges. When comparing this with the HN model, the question arises regarding which state serves as the counterpart of the steady state in the HN model.

One potential approach is to consider the right eigenvector $|\varphi_j^R\rangle$ with the largest imaginary part of its eigenvalue, i.e., $\max\left[\mathrm{Im}(\nu_j)\right]$. This vector would represent the steady state of $\hat H_\mathrm{HN}$, provided that we renormalize it under time-evolution. However, for open boundary conditions, the spectrum is real. Instead, we consider the eigenvector with the smallest eigenvalue, which corresponds to the state $|\varphi_N^R\rangle$ as defined in Eq.~(\ref{eigvec1}). In this way, we argue that this state, in some sense, mimics the ground state of the system. The fidelity between this state and the steady state of the LME is defined as
\begin{equation}\label{fid}
    F=\langle\varphi_N^R|\hat\rho_\mathrm{ss}|\varphi_N^R\rangle,
\end{equation}

The spectrum and eigenvectors of the LME are determined through exact diagonalization of the vectorized Liouvillian~(\ref{lindbladvec}). The numerical results for the three defined quantities are presented in Fig.~\ref{fig2}. In the upper plot (a), we display the position~(\ref{pos}) for both the LME and the HN model. We vary the rate $\gamma$ and consider two system sizes. In the lower plot (b), we provide the purity~(\ref{pur}) of the steady state and the fidelity~(\ref{fid}). There is a significant distinction between the two models. The HN model exhibits edge localization, or the skin effect, for all $\gamma$ values, and for $\gamma=1$, only the rightmost site is populated.

In contrast, the LME supports two different phases: a delocalized phase for $\gamma<\gamma_c=1/2$ and a localized phase for $\gamma>\gamma_c$. In the localized phase, the state primarily occupies the right edge, resembling a skin state, while in the delocalized phase, the state spreads across the entire lattice. The purity, as seen in (b), illustrates that the delocalized phase is approximately a maximally mixed state. Although the states $\hat\rho_\mathrm{ss}$ and $|\varphi_N^R\rangle$ exhibit similar localization properties near $\gamma=1$, the fidelity reveals that they are, in fact, quite distinct.

Even though only the steady state represents a physical state, we can still analyze the properties of the remaining eigenvectors of the Liouvillian. Not shown here, but for $\gamma>\gamma_c$, they tend to localize at the edge, while for $\gamma<\gamma_c$, they become delocalized.

\begin{figure}[h]
\centerline{\includegraphics[width=8.4cm]{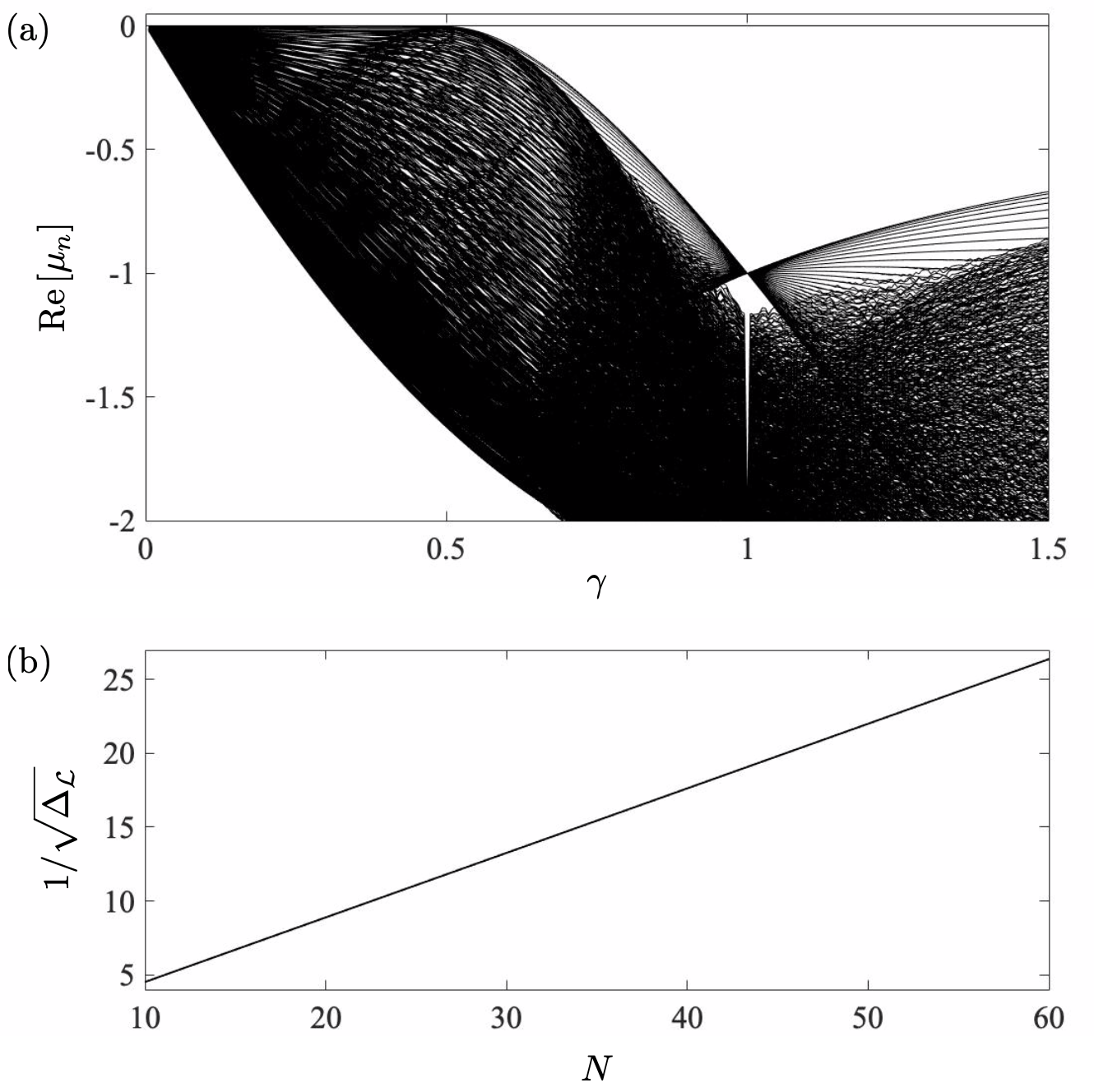}}
\caption{Real part of the Liouvillian spectrum $\mu_j$ (a) as a function of $\gamma$ for a fixed system size $N=41$ (a). The lower plot (b) demonstrate the gap closening ($\Delta_\mathcal{L}\rightarrow0$ as $N\rightarrow\infty$) at the critical point $\gamma=1/2$. It is evident from the upper plot how the Liouvillian gap, $\Delta_\mathcal{L}$, opens up beyond $\gamma=1/2$, while for $\gamma<1/2$, the numerical results indicates a gapless, continuous spectrum. From (b), we find the power-law scaling with an exponent $\nu=1/2$ characteristic of a for a mean-field critical point. } \label{fig3}
\end{figure}

To provide insight into the critical properties of the delocalized-to-localized phase transition, we plot the real part of the Liouvillian spectrum $\mu_j$ in Fig.~\ref{fig3} (a). It is evident that the Liouvillian gap $\Delta_\mathcal{L}$, as defined in Eq.~(\ref{lgap}), closes precisely at the critical point $\gamma_c$. In this figure, we consider a lattice with $N=41$ sites, resulting in a Liouvillian matrix of dimensions $1681\times1681$. In the thermodynamic limit, the real part of the spectrum becomes gapless and continuous within the delocalized phase.

For a critical model, the closure of the gap concerning the system size typically follows universal scaling:
\begin{equation}
    \Delta_\mathcal{L}\sim N^{-1/\nu},
\end{equation}
where $\nu$ represents the correlation length critical exponent~\cite{cardy}. Our numerical findings affirm that the exponent $\nu$ is 1/2, as demonstrated in Fig.~\ref{fig2} (b). This value corresponds to the exponent of a mean-field critical model, suggesting that quantum fluctuations are suppressed in comparison to fluctuations arising from the environment.

It might appear counterintuitive that the ``fluctuation-induced breakdown'' of the skin effect occurs for weak couplings ($\gamma$) rather than when the system is strongly coupled to its environment. This can be understood by considering the steady state, which describes the system after an infinitely long time. In the delocalized phase, the system is gapless, and the relaxation time diverges. Consequently, there is more room for fluctuations to manifest compared to the localized phase. This phenomenon resonates with what is known in adiabatic quantum computing and quantum control in atomic physics~\cite{stirap}.

For reasons that will become clear, let us introduce a generalized Fock space. The Liouvillian Fock states, denoted as $|l\rangle\!\rangle$ and residing in the Liouvillian space, can be identified through vectorization as
\begin{equation}
    |n\rangle\langle m|\rightarrow|N(n-1)+m\rangle\!\rangle,
\end{equation}
such that $1\leq l\leq N^2$. The vectorized LME is provided in~(\ref{lindbladvec}), and its formal solution is represented as
\begin{equation}
    |\rho(t)\rangle\!\rangle=\exp\left(\hat{\mathcal{L}}_vt\right)|\rho(0)\rangle\!\rangle
\end{equation}. 
We can now extend the concept of Fock state lattices~\cite{pil} to Liouvillian Fock state lattices. The idea here is to envision the Liouvillian matrix $\hat{\mathcal{L}}_v$ as describing hopping in a lattice, with its sites representing the Fock states $|l\rangle\!\rangle$. In other words, the components of the vector $|\rho\rangle\!\rangle$ correspond to the populations at different lattice sites. Specifically, the diagonal elements $\varepsilon_i$, defined as $\varepsilon_i\equiv\langle\!\langle l|\hat{\mathcal{L}}_v|l\rangle\!\rangle$, represent on-site ``energies'', while the off-diagonal elements $\langle\!\langle l|\hat{\mathcal{L}}_v|k\rangle\!\rangle$ characterize the tunneling amplitudes between sites $l$ and $k$. Importantly, $\varepsilon_l$ is real and $\varepsilon_l\leq0$, implying that the diagonal terms induce on-site dissipation.

\begin{figure}[h]
\centerline{\includegraphics[width=5cm]{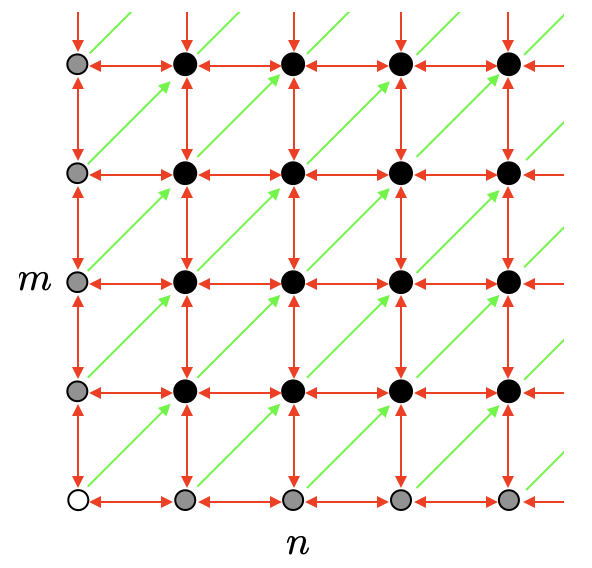}}
\caption{The Liouvillian Fock state lattice. Each lattice point represents a Liouvillian Fock state $|l\rangle\!\rangle\leftrightarrow|n\rangle\langle m|$, and the colors denote the onsite (imaginary) ``energy'' with white ($=0$) $>$ gray $>$ black. Consequently, apart from the lower left corner site, the ``population'' within the other sites decays are suppressed by dissipation. The arrows describe the non-vanishing tunneling elements. If quantum jumps, i.e., fluctuations, are disregarded, there would be no diagonal tunneling (green arrows between next-nearest neighbors). The red arrows alone do not have a preferred direction. The onsite dissipation tend to localize the state to the lower left corner, while the tunneling terms counteract the localization. For the HN model, one finds the same square lattice, but there is no diagonal tunneling, and the horizontal/vertical tunnelings are imbalanced with amplitudes $1\pm\gamma$, as expected for the HN model.} \label{fig4}
\end{figure}

In the context of the problem at hand, it turns out that the Liouvillian Fock state lattice takes the form of a square lattice, as illustrated in Fig.~\ref{fig4}. The thick dots represent the lattice sites, and the arrows indicate the allowed tunnelings between these sites. The color shading of the sites reflects the magnitude of on-site dissipation, with a decreasing order from black to gray to white.

Having identified the Fock state lattice, we make the following observations. There is {\bf no} asymmetry in the horizontal/vertical tunnelings, as one might expect from an HN model. However, there is diagonal tunneling that is non-zero only in one direction. In addition to the imaginary on-site terms, these diagonal directional tunneling terms imply that the Liouvillian matrix becomes NH. Furthermore, these diagonal processes result from quantum jumps, driven by fluctuations, and consequently, they do not appear in the lattice emerging from the HN model. More precisely, the corresponding HN lattice is simply a two-dimensional version of the HN model with imbalanced left/right tunnelings. Hence, quantum jumps not only induce diagonal tunneling terms but also alter the nearest neighbor tunneling amplitudes, making them balanced.

As previously mentioned, the on-site ``energies'' vary in the lattice. For instance, at site $l=N^2$ (corresponding to the rightmost lattice site in the original 1D real space lattice), we have $\varepsilon_{N^2}=0$, indicating no local dissipation. In the Fock state lattice, this corresponds to the site in the lower left corner (white dot). Along the edges originating from the $l=N^2$-site, there is moderate dissipation (gray dots), while in the bulk, the dissipation is most pronounced (black dots). Let us introduce $\varepsilon_b$ and $\varepsilon_e$ for the bulk and edge on-site ``energies'' respectively (the black and gray sites in the figure), and $t_0$ and $t_d$ for the horizontal/vertical (nearest neighbor) and diagonal (next nearest neighbor) tunneling amplitudes respectively. These lattice parameters are related to $\gamma$ as shown in Tab.~\ref{tab1}. If there were no tunnelings ($t_0=t_d=0$), the system (ground/steady state) would localize at the site represented by the white dot. The tunneling terms tend to delocalize the steady state. As $\gamma$ increases, the dissipation-induced localization becomes stronger, but at the same time, tunneling-induced delocalization also strengthens. Before the critical point, $\gamma<\gamma_c=1/2$, the nearest neighbor tunneling rates ($2|t_0|$) are greater than $\varepsilon_b$, causing the system to delocalize. Beyond the critical point, $\gamma>\gamma_c$, where $2|t_0|<\varepsilon_b$, the edge skin mode wins. This reasoning demonstrates how the transition can be understood from the Liouvillian Fock state lattice.

\begin{table}[h]
\begin{tabular}{ |c|c|  }
\hline
$\hspace{0.2cm}$Nearest neighbour tunneling$\hspace{0.2cm}$   & $\hspace{0.2cm}t_0=\pm i(1-\gamma)\hspace{0.2cm}$\\
 \hline
Diagonal tunneling &   $t_d=2\gamma$\\
 \hline
 Bulk ``energy'' & $\varepsilon_b=-2\gamma$\\
 \hline
 Edge ``energy'' & $\varepsilon_e=-\gamma$\\
 \hline
 
\end{tabular}
\caption{Table of the parameters forming the Fock state lattice of Fig.~\ref{fig4}, i.e. green arrows -- $t_d$, red arrows -- $t_0$, black dots (bulk sites) -- $\varepsilon_b$, and gray dots (edge sites) -- $\varepsilon_e$. Not shown in the figure are the upper and right edges, where onsite ``energies'' are set to $-3\gamma$. For sufficiently large lattices, these edge sites should not significantly impact the criticality.
}\label{tab1}
\end{table}


It is worth noting that similar delocalization-localization transitions have been discussed in the past with a model described as
\begin{equation}\label{bismod}
    \frac{\partial\hat\rho}{\partial t}=i\left[\hat\rho,\hat S_x\right]+\frac{\gamma}{S}\left(2\hat S^-\hat\rho\hat S^+-\hat S^+\hat S^-\hat\rho-\hat\rho\hat S^+\hat S^-\right).
\end{equation}
This model was introduced in the late 1970s to study quantum optical bistability~\cite{optbis}. In this context, the $\hat S$-operators represent collective spin operators, and $S$ is the total (conserved) spin. The Hamiltonian can be seen as describing a classical drive of the spin, while the Lindblad dissipation represents spontaneous decay of the spin. One attractive aspect of this model is that the steady state can be determined analytically~\cite{optbis2}, and it exhibits a mean-field critical point at $\gamma_c=1/2$, where the system transitions from being magnetized (localized) to paramagnetic (delocalized)~\cite{optbiscrit2}. The transition is continuous, without any apparent spontaneous symmetry breaking, which has generated some debate~\cite{optbiscrit2}.

To draw a connection to the model studied in this paper, it is important to note that the Hamiltonian component $\hat S_x$ tends to delocalize the state in the spin Fock basis (the $|S,m\rangle$-eigenstates of $\hat S_z$), while the dissipative part drives the state toward the ``edge'' $|S,-m\rangle$. We have numerically solved the LME using $\hat H=\hat E_x$ and $\hat L=\hat E$, with $\hat E_x$ and $\hat E$ being operators from the Euclidean algebra~(\ref{ealg}). This resulted in a similar phase transition as observed for the LME~(\ref{collind}). Consequently, we draw a comparison between two models: one supporting an $SU(2)$ algebra and the other an Euclidean algebra, both being otherwise equivalent. Of course, the spin operators come with a square-root normalization factor when acting on the Fock states, but this factor does not alter the Fock state lattice geometry; it induces a strain in the lattice~\cite{pil}. It is important to mention that this connection is relevant only for a finite lattice, as in the infinite case, all operators would mutually commute, i.e. $\left[\hat E,\hat E^\dagger\right]=\left[\hat E,\hat E_x\right]=\left[\hat E^\dagger,\hat E_x\right]=0$.

The critical behavior of the bistability model has been extensively studied in Ref.~\cite{optbiscrit2}. It was argued that the critical behavior, devoid of symmetry breaking, arises from the transition `softening' of a first-order transition into a continuous one. In the current model, we appear to observe similar universal behavior. Specifically, the transition is continuous, the steady state remains unique throughout, and hence, there is no apparent symmetry breaking occurring. It is worth noting that quenched disorder can alter the nature of a transition from first to second order in classical critical models~\cite{qd}. Tri-critical points provide another example where the order of a transition changes, and in the Potts model, the transition can shift from first to second order as a system parameter varies~\cite{potts}. Continuous phase transitions without symmetry breaking can also occur in fermionic models when the Fermi sea undergoes volume changes~\cite{fermisea}.

It remains unclear whether the mechanism underlying the observed criticality in this open system (and in the bistability model of  Eq.~(\ref{bismod})) differs in nature from those listed in the references mentioned above. After all, we are dealing with an open, non-Hamiltonian system. In Ref.~\cite{optbiscrit2}, it was noted that the full Hamiltonian model, including the degrees of freedom of the environment, exhibited a first-order phase transition, and it was only in the limiting case of infinite separation in time-scales that the initially first-order, discontinuous phase transition became continuous. Something similar might occur in our system, such as starting from the Hamiltonian~(\ref{fullham}) and coupling it to a bath of oscillators.

\subsection{Non-hermitian sensors}\label{sensorsec}
In recent years, there have been several proposals on how systems described by NH ``Hamiltonians'' can be leveraged to enhance sensor performance. Various concepts for these implementations have been explored, including harnessing the non-analyticity associated with exceptional points~\cite{sensor3}, non-unitary evolution~\cite{sensor2}, and the phenomenon of exponential sensitivity~\cite{sensor}. Motivated in part by a recent experimental demonstration~\cite{sensor4}, our focus will be on the latter aspect and its application to the HN model.

We aim to address two key questions in this context: firstly, what role do fluctuations play in the sensor setup, and secondly, how does disorder impact the performance of the NH sensor?

\begin{figure}[h]
\centerline{\includegraphics[width=8cm]{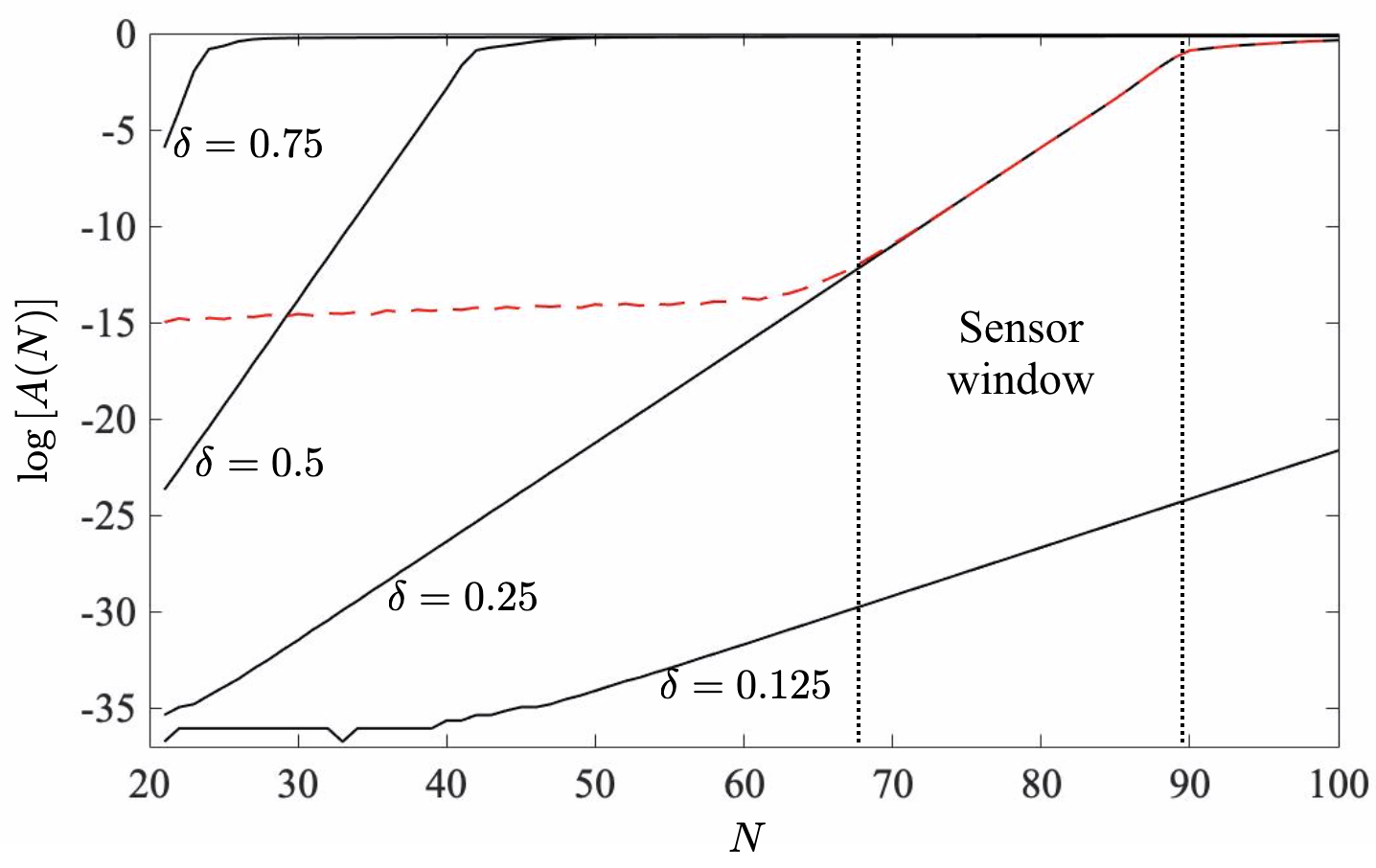}}
\caption{Logarithm of the HN auto-correlation Function~(\ref{auto}), shown as a function of the system size $N$ for four different asymmetry parameters $\delta$ (solid black lines). In all four examples, the perturbation strength is $\epsilon=10^{-10}$. The linearity of the curves clearly demonstrates the exponential $N$-dependence and the point at which the perturbation $\delta\nu$ transitions from being real to complex, indicated as a ``knee'' in the curves. The dashed red line represents an example of the disordered HN sensor (averaged over 1000 disorder realizations) with a disorder strength $W=0.0005$ (i.e., $0.05\%$ of the tunneling rate). In the case of disorder, the sensor must be of sufficient size for the perturbation to outweigh the effects of random disorder. In practice, there exists a window, as marked for the $\delta=0.25$ curve in the plot, within which the sensor operates effectively (see also Fig.~\ref{fig7}). For a sensor that is too small, the perturbation is overshadowed by disorder, and for a sensor that is too large, the perturbation becomes complex.  } \label{fig5}
\end{figure}

\subsubsection{Non-hermitian sensor with fluctuations}
Let us begin by summarizing the fundamental concept of the NH sensor~\cite{sensor}. As previously mentioned, the skin effect is directly linked to exponential sensitivity. To illustrate this, let us consider the $N$-site open BC HN model. The spectrum for this model was given in Eq.~(\ref{obc}), and we observed how it is real and symmetric around zero. This symmetry implies that for an odd number of sites, there is a zero eigenvalue, denoted as $\nu_z\equiv\nu_j=0$ for $j=(N+1)/2$. Now, let us further assume a very weak (real) coupling denoted as $|\epsilon|\ll1$, which connects the first and last sites. This coupling represents a non-local perturbation of the form: 
\begin{equation}\label{pert}
    \hat V=\epsilon\left(\hat a_1^\dagger\hat a_N+h.c.\right).
\end{equation}
Our goal here is to determine the value of $|\epsilon|$. When treating $\hat V$ as a perturbation, we find (for NH, perturbation theory involves both left and right unperturbed states since they form the orthogonal states used for the resolution of identity) that the lowest-order corrections to the zeroth eigenvalue scale as~\cite{sensor}
\begin{equation}\label{expsens}
    \nu_z\rightarrow\nu_z=\delta\nu,\hspace{0.7cm}\delta\nu\sim\epsilon e^{\alpha N}.
\end{equation}
Here, $\alpha$ depends on specific system details. Consequently, the eigenvalue remains real but shifts away from zero. Notably, this shift can be significant as long as $N$ is sufficiently large. Beyond a critical perturbation $\epsilon_c(N)$ (which depends on the system size and other system parameters), the scaling breaks down, and the eigenvalue becomes complex. In line with the experimental work of Ref.~\cite{sensor4}, if the system is initially prepared in the zero eigenvalue state $|\varphi_z^R\rangle$ and evolves for a brief time $t$ under the perturbed Hamiltonian $\hat H_\mathrm{HN}+\hat V$, resulting in the state $|\psi\rangle$, then the decay of the auto-correlation function
\begin{equation}\label{auto}
    A(N)=|\langle\varphi_z^L|\psi\rangle|
\end{equation}
serves as a measure of the perturbation. 

Consequently, for a given perturbation strength $\epsilon$, due to the exponential sensitivity, the logarithm of $A(N)$ should exhibit a linear relationship with the system size N as long as $\epsilon<\epsilon_c(N)$. The results, obtained from numerical time-propagation of the initial state, are presented in Fig.~\ref{fig5} as solid black curves. In the figure, we display the logarithm of the auto-correlation function for four different HN parameters, denoted as $\delta$. It is worth recalling that in the Lindblad realization of the HN model, $\delta$ corresponds to the rate $\gamma$. As expected, the anticipated linear dependence on $N$ is demonstrated in all four examples within the figure.

It is particularly noteworthy that for large $\delta$ values, the exponential dependence as per Eq.~(\ref{expsens}) persists only for relatively small system sizes. Conversely, this regime can extend over significantly larger system sizes for small $\delta$ values. The slope of the curves is determined by the parameter $\alpha$ in Eq.~(\ref{expsens}), which is found to scale as $\alpha\sim\delta$. This relationship explains why rather large system sizes $N$ are required to reach the critical $\epsilon_c$ for small $\delta$ values. In practical implementations, state tomography provides the time-evolved state $|\psi(t)\rangle$. With prior knowledge of $|\varphi_z^L\rangle$, it becomes possible to estimate the decay of $A(N)$ and, consequently, $\delta\nu$. This approach has been realized in classical light-pulses in waveguides, where fluctuations likely play a less significant role (see discussions in Sec.~\ref{sec:con})~\cite{sensor4}.

\begin{figure}[h]
\centerline{\includegraphics[width=8cm]{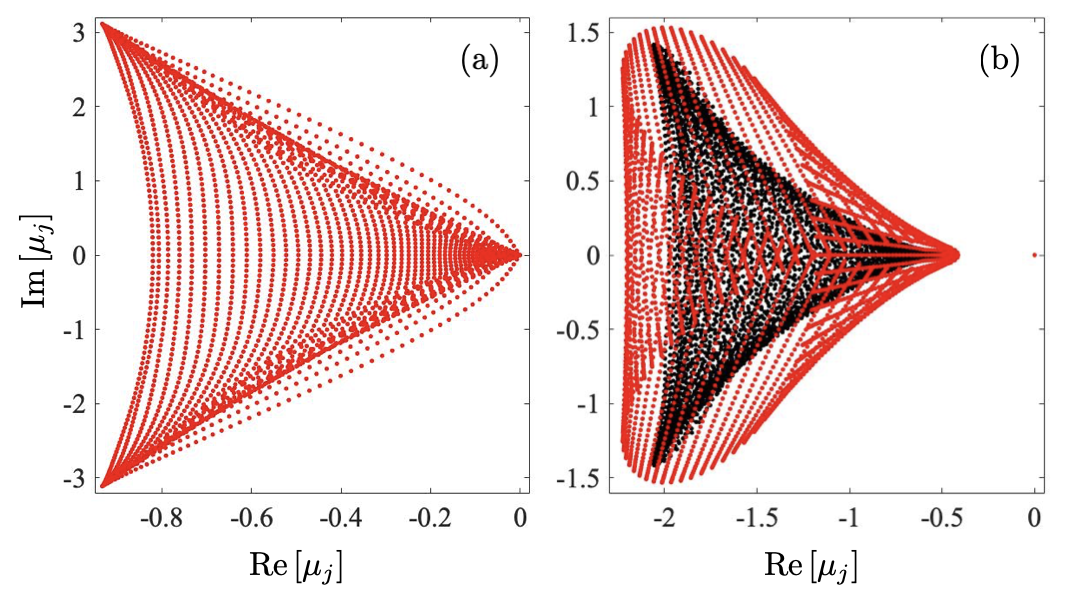}}
\caption{The spectrum of the unperturbed (black dots) and perturbed Liouvillian (red dots) for the localized phase (a) and the delocalized phase (b). In the delocalized phase of (a) with $\delta=0.25$, it is challenging to distinguish between red and black dots. However, in the localized phase (b) with $\delta=0.75$, there is a clear exponential sensitivity. These results are based on a system size of $N=61$ and a perturbation strength of $\epsilon=10^{-10}$. } \label{fig6}
\end{figure}

Now, turning our attention to the LME, we consider the same type of perturbation~(\ref{pert}). The exponential sensitivity in this context is a consequence of the skin effect. As we have discussed, the skin effect can be lost in the LME, leading to the steady state becoming delocalized when $\gamma<\gamma_c=1/2$. Consequently, we expect that exponential sensitivity is only present in the localized phase. Indeed, within this phase, the eigenvectors of the Liouvillian become localized to the corner of the Liouvillian Fock state lattice~\cite{com2}. In Fig.~\ref{fig6}, we provide two examples of the Liouvillian spectrum: one in the delocalized phase (a) and the other in the localized phase (b). As anticipated, the exponential sensitivity is no longer present when the system transitions into the delocalized phase.

However, in the localized phase, the Liouvillian also exhibits exponential sensitivity, a phenomenon that has been observed in the past~\cite{nhlindblad1,nhlindblad2,nhlindblad2b,sofia}. Nevertheless, the challenge remains to identify an observable quantity that is exponentially sensitive and capable of extracting information about the perturbation.
 
For the HN model, a natural choice is to consider the state associated with the zero eigenvalue, $|\varphi_z^R\rangle\!\rangle$. As we have seen, this is due to the initial decay of its auto-correlation function, which directly reflects the magnitude of the perturbation. In the case of the full LME, the primary option would be the steady state, denoted as $\hat\rho_\mathrm{ss}^{(\epsilon=0)}$ (where $\epsilon=0$ signifies the unperturbed state). We can analyze how this steady state evolves under the perturbed Liouvillian. Alternatively, we could also contemplate using the same initial state $|\varphi_z^R\rangle$ as in the HN model and explore the influence of fluctuations on the auto-correlation function~(\ref{auto}). Through numerical simulations, we find no signs of exponential sensitivity in either of these scenarios.

In the latter case, the decay of the auto-correlation function is predominantly driven by the relaxation towards the system's steady state, and it occurs on a timescale proportional to the inverse Liouvillian gap~(\ref{lgap}). The perturbation-induced decay happens on an entirely different timescale and gets obscured by the relaxation of the steady state. Extending the steady state relaxation time by moving closer to the critical point does not result in a favorable situation, as it rapidly suppresses exponential sensitivity when the system becomes more delocalized. In our numerical experiments, we have not identified a regime where the system effectively functions as a sensor when initialized with the $|\varphi_z^R\rangle$ state.

This same argument applies to initializing the system in the unperturbed steady state, meaning that the relaxation of $\hat\rho_\mathrm{ss}^{(\epsilon=0)}$ to the final perturbed steady state $\hat\rho_\mathrm{ss}^{(\epsilon)}$ dominates the evolution. Importantly, this behavior does not exhibit a strong $N$-dependence. More precisely, we find that the auto-correlation function $A(N)=\mathrm{Tr}\left[\hat\rho_\mathrm{ss}^{(0)}\hat\rho_\mathrm{ss}^{(\epsilon)}\right]$ follows a linear $N$-dependence rather than an exponential one.

In summary, even though the spectrum of the Liouvillian exhibits exponential sensitivity to non-local perturbations in the localized phase, it is not evident how this can be effectively harnessed for sensing purposes. Generalizing the concept of the HN sensor does not appear to yield favorable results. It remains uncertain whether there might be another experimentally measurable quantity that could salvage the sensor's performance when accounting for fluctuations. 

\subsubsection{Non-hermitian sensor with disorder}
Another significant limitation of the NH sensor that we need to address, which does not stem from environment-induced fluctuations, pertains to disorder. Consider the presence of local quenched disorder, which can arise in an imperfect sensor, causing the actual Hamiltonian to take the form:
\begin{equation}
    \hat H_\mathrm{dHN}=\hat H_\mathrm{HN}+\sum_{n=1}^N\kappa_n\hat n_n,
\end{equation}
where $\kappa_n\in\left[-W,W\right]$ represents a random onsite offset. Here, $W$ denotes the disorder strength. We assume that $W\ll1$, which means it is orders of magnitude smaller than the tunneling rate, ensuring that the system is not localized on any relevant length scales. However, it is important to note that $W$ can be larger than the perturbation strength that the sensor is meant to measure. Therefore, a potential breakdown of the sensor should not be attributed to hindrance in propagation due to localization.

\begin{figure}[h]
\centerline{\includegraphics[width=7cm]{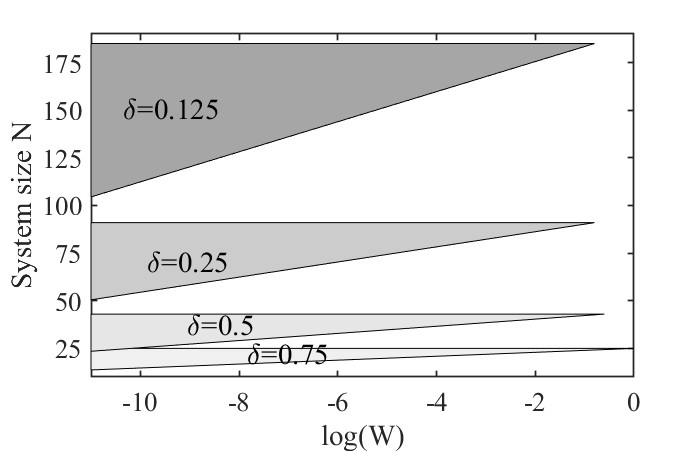}}
\caption{Gray regions indicate the parameter regimes in the $NW$-plane (note the logarithmic scale of the disorder strength $W$) where the sensor can effectively operate. Above this region, the $N$-dependence of the perturbation is no longer exponential, and below, disorder-induced noise dominates the detector signal. The perturbation strength is kept the same as in Figure 5, i.e., $\epsilon=10^{-10}$. } \label{fig7}
\end{figure}

In Fig.~\ref{fig5}, we present the sensor's performance in the presence of quenched disorder, indicated by the red dashed line. The disorder strength is approximately $0.05\%$ of the tunneling strength, and to reduce the scatter, we averaged over 1000 disorder realizations. Our findings reveal that below a certain system size, denoted as $N_l$, the sensor primarily detects the disorder, with the perturbation signal getting lost in the noise generated by the disorder. However, beyond $N_l$, the perturbation signal begins to dominate the disorder noise, thanks to its exponential increase, and the disorder's impact on the signal diminishes significantly.

It is important to note that above a certain upper system size, $N_u$, the perturbation signal no longer follows the exponential form described in Eq.~(\ref{expsens}). Consequently, when given a perturbation $\epsilon$ and a disorder strength $W$, there exists a window of system sizes $N_l<N<N_u$ within which the NH sensor operates effectively. These windows, corresponding to the same values of $\delta$ as seen in Fig.~\ref{fig5}, are displayed in Fig.~\ref{fig7}.

In summary, when $\delta$ is small, it implies the need for larger sensor sizes $N$. In practice, achieving extremely large chains may not be feasible, making it desirable to use a larger $\delta$. Our results are computed using the same methodology as for the non-disordered sensor, involving the propagation of the initial state and the analysis of the auto-correlation function's decay. We have also verified numerically that similar results can be obtained by directly extracting $\delta\nu$ from the full spectrum, rather than indirectly calculating it from $A(N)$.

\section{Discussion and Concluding remarks}\label{sec:con}
While NH Hamiltonians offer utility in modeling a quantum system's interaction with its environment, they can also lead to complex, non-physical outcomes. A significant challenge in this endeavor arises from the intricate entanglement between the quantum system and its environment, resulting in a mixed state representation rather than the pure state representation $\hat\rho=|\psi\rangle\langle\psi|$. However, it is worth noting that there are scenarios in which the quantum system's state remains nearly pure. In such cases, relevant observables $\mathcal{O}=\mathrm{Tr}\left[\hat\rho\hat O\right]$ can be well described by a pure state. Many experimental activities involve classical emulations of quantum systems, where classical systems obey equations of motion similar, or equivalent, to those found in quantum systems described by NH Hamiltonians. While classical systems are not represented by states in a Hilbert space, the fluctuation theorem extends to both classical and quantum systems. In a strict sense, any open system is subject to fluctuations from its environment, with Brownian motion serving as a classic example. However, this influence may be negligible for macroscopic objects. Classical states, described as coherent states in the quantum realm, are known to be robust against fluctuations, such as the state of the electromagnetic field originating from a laser~\cite{com3}.

The role of fluctuations becomes a more intricate matter when dealing with systems deep in the quantum regime. To circumvent this issue, researchers have employed the concept of post-selection~\cite{postselect}, where experiments are conducted under full observation, and only the data from experimental runs that do not undergo ``quantum jumps'', such as spontaneous photon emission, are considered~\cite{unravel}. However, this approach often leads to a significantly reduced probability of successful experimental runs over time, as they become exponentially suppressed.

In this study, we explored the time evolution generated by the Liouvillian without invoking measurement-induced projections or assuming a semi-classical regime. Specifically, we focused on two well-studied phenomena within the framework of NH QM: the skin effect and NH sensors.

While previous arguments suggested that both the skin effect and NH sensors should persist when fluctuations are taken into account~\cite{nhlindblad1,nhlindblad2,nhlindblad4}, our study unveiled subtleties in this context, leading to new insights. Importantly, the Liouvillian is not uniquely determined by a NH ``Hamiltonian''. This multiplicity is akin to the purification of mixed states~\cite{nc}, where infinitely many different pure states can construct a given mixed state. Consequently, there exist infinitely many Liouvillians corresponding to equivalent NH ``Hamiltonians'', and these Liouvillians can exhibit qualitative differences.

In our study, we chose to investigate a Liouvillian represented by collective quantum jumps, as defined in Eq.~(\ref{collind}). We also considered an alternative model with local jump operators, as defined in Eq.~(\ref{locjump}). Upon numerical analysis of the latter model, we found that it lacks critical behavior – the delocalized phase does not emerge in this case. This finding implies that the collectiveness of quantum jumps is essential for the appearance of a critical point. The criticality observed in our model is an example of a driven-dissipative non-equilibrium phase transition~\cite{lcrit}. This specific type of transition is qualitatively different from phase transitions described by the Ginzburg-Landau paradigm~\cite{cardy}. The transition is continuous and lacks any apparent symmetry breaking, yet the critical exponent for the correlation length aligns with the characteristics of a mean-field transition. This observation resonates with related models, such as the one represented in Eq.~(\ref{bismod}), which has been explored in the context of optical bistability.
  
The existence of a delocalized phase signifies the breakdown of the NH skin effect. In this phase, the system exhibits an extended steady state that approximately populates the lattice sites uniformly. The coherences between the sites practically vanish, implying that the steady state approximates a maximally mixed state or an infinite-temperature state. In the case of periodic BC, the maximally mixed state is the exact steady state, a concept that was briefly mentioned in a previous study~\cite{hatanonelsonlindblad}, which investigated a quadratic Liouvillian.

It is essential to recognize that not only the steady state becomes delocalized for $\gamma<\gamma_c$, but the other Liouvillian eigenvectors $\hat\rho_j$ also undergo delocalization in this phase. This delocalization affects the entire spectrum of the Liouvillian.

Indeed, the absence of a skin effect in the delocalized phase results in the system's lack of exponential sensitivity. Consequently, it cannot be employed for sensing purposes in this regime. However, when the losses are substantial, i.e., $\gamma>\gamma_c$, the Liouvillian does exhibit exponential sensitivity once it enters the localized phase.

Nonetheless, despite the presence of exponential sensitivity in this regime, our exploration did not yield a suitable measure that could effectively extract the quantity to be detected. Attempts to directly generalize experiments like the one reported in~\cite{sensor4}, which used classical light, were ineffective in the context of Liouvillian systems. This inefficacy is primarily due to the dominance of other mechanisms, such as relaxation towards a steady state, during the early stages of evolution. We also considered observing the long-term evolution and the fidelities of the resulting steady states, but these did not exhibit exponential sensitivity either.

While the specific observables explored in this study did not yield the desired results, it remains a possibility that more refined observables could be identified for use in Liouvillian-based sensors. However, further investigation and research would be required to identify and assess these potential observables effectively.
 
In our study, we have also demonstrated that disorder introduces limitations to the applicability of NH sensors, even without considering fluctuations. This finding underscores the importance of the sensor's size relative to the disorder strength. The sensor must be sufficiently large to overcome the noise due to disorder.

In particular, we observed that for the NH sensor to effectively detect a perturbation signal, the signal strength must exceed a critical value that is determined by the disorder strength. This means that the sensor must be sufficiently large to overcome the detector noise generated by the disorder.

These findings emphasize the importance of carefully considering disorder effects and ensuring that the sensor size and sensitivity are suitable for practical applications. It also suggests that engineering sensors with greater robustness against disorder may be necessary for reliable measurements in realistic experimental conditions.

While our analysis has been centered on the HN model, the applicability of our results to other models remains an open question. In Appendix~\ref{sec:app2}, however, we extend our findings to the NH SSH chain and demonstrate that the qualitative conclusions applies. Nevertheless, one can imagine the potential for further research in other directions that addresses slightly different questions, including topology, localization, and the realm of many-body NH physics. We propose that the introduction of the Liouvillian Fock state lattice, as illustrated in Fig.~\ref{fig4}, offers a promising tool for gaining fresh insights. Particularly fascinating is the observation that, for the 1D HN model, the dimension of the Liouvillian Fock state lattice stands at 2D. This observation leads to the intriguing speculation that, for a D-dimensional model coupled to an environment, the relevant physics might manifest in D+1 dimensions. Furthermore, the interplay of symmetries in NH quantum mechanics compared to the full LME~\cite{lsym} remains an open question, partially due to the complexity of relating biorthogonal QM to real physical systems. These questions are left unexplored for future studies.

\appendix

\section{Open quantum systems}\label{sec:app1}

\begin{figure}[h]
\centerline{\includegraphics[width=5cm]{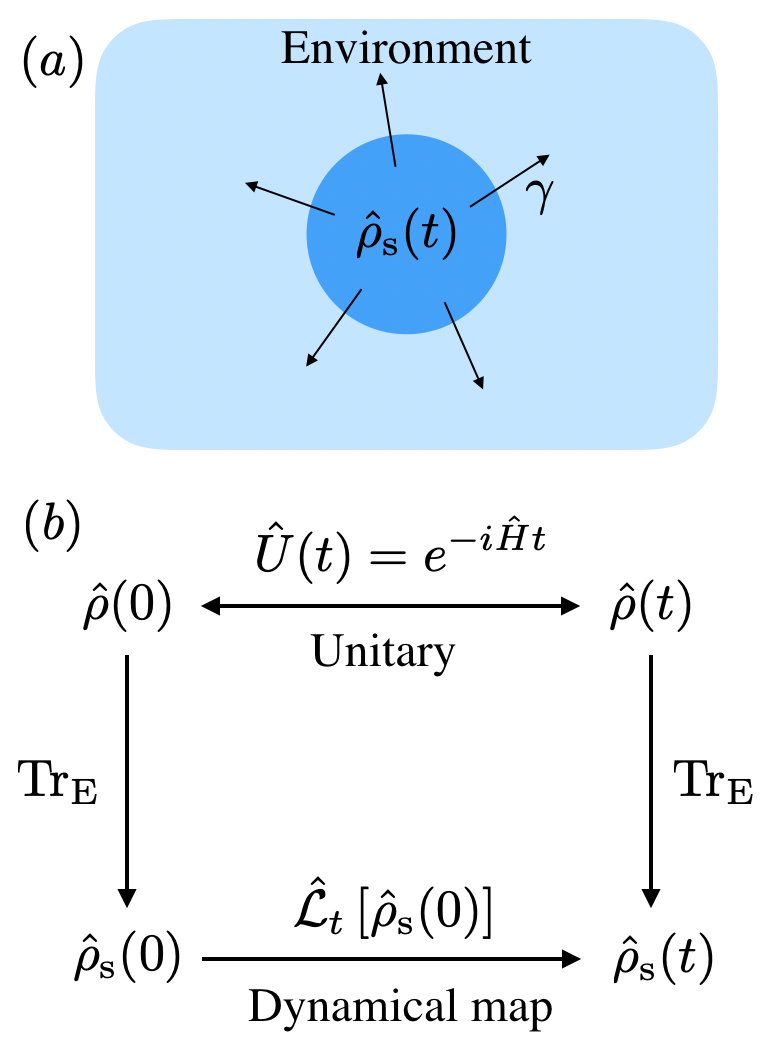}}
\caption{Description of the open system evolution. The (sub)system $\mathcal S$ is represented by a state $\hat\rho_\mathrm{s}(t)$, schematically depicted in blue in (a). The system interacts weakly with an environment $\mathcal E$ (light blue), allowing information within $\mathcal S$ to irreversibly dissipate into $\mathcal E$ with a characteristic decay rate $\gamma$. In (b), the concept of the CPTP dynamical map is presented. The full system comprises $\mathcal S$ and $\mathcal E$, and its state $\hat\rho(t)$ undergoes unitary evolution under a Hamiltonian $\hat H$, as indicated by the upper horizontal double arrow. The system state $\hat\rho_\mathrm{s}(t)$ is derived by taking the partial trace of the full state over the environment degrees of freedom, represented by vertical downward arrows. The dynamical map $\hat\rho_\mathrm{s}(t)$, shown as the lower horizontal arrow, governs the evolution of the system state, transforming $\hat\rho_\mathrm{s}(0)$ into $\hat\rho_\mathrm{s}(t)$. While the full state evolution is unitary and thus reversible, all other steps (partial traces and dynamical map) are irreversible. } \label{figa1}
\end{figure}

\begin{figure}[h]
\centerline{\includegraphics[width=8cm]{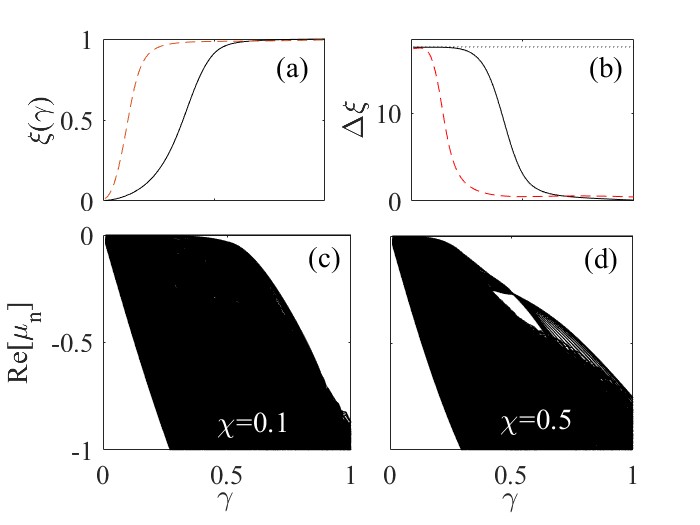}}
\caption{Demonstration of the localized-delocalized transition for the NH SSH model. In (a), we present the same information as Fig.~\ref{fig2} (a) for the HN model. For large $\gamma$ values, the steady state becomes localized at the edge of the chain, whereas for weak $\gamma$ values, it becomes delocalized over the entire lattice, as evident in (b), which displays the width~(\ref{wid}). The dashed red curves correspond to $\chi=0.5$, and the solid black curves to $\chi=0.1$. The dotted line in (b) represents the width for a maximally delocalized state. It is evident that the SSH parameter $\chi$ smoothens the transitions, but criticality appears to persist. This is further confirmed in (c) and (d) for the real parts of the Liouvillian spectra. } \label{figa2}
\end{figure}

In this appendix, we outline some general concepts related to the dynamics of open quantum systems. The typical scenario is illustrated in Fig.~\ref{fig1}. In (a), we depict a small system $\mathcal{S}$ interacting with a larger environment $\mathcal{E}$. Information is exchanged between these subsystems, with the rate of exchange determined by the parameter $\gamma$. This exchange includes processes such as particle or energy losses and decoherence. The combined system's evolution is described by the Schrödinger equation. For instance, an initial pure separable state evolves as
\begin{equation}
|\Psi(t)\rangle=\hat U(t)|\psi_\mathrm{s}(0)\rangle\otimes|\psi_\mathrm{E}(0)\rangle.
\end{equation}
Here, the time-evolution operator $\hat U(t)$ is generated from the full Hamiltonian $\hat H=\hat H_\mathrm{s}+\hat H_\mathrm{E}+\hat H_\mathrm{sE}$, where the first two terms represent the system and environment sub-Hamiltonians, and the last term accounts for their interaction. The state of the system at time $t$ is obtained by taking the partial trace of the full state, as illustrated in Fig.~\ref{figa2} (b).
 
As time progresses, the initially separable state of the system-environment becomes entangled. Given that the initial state is pure, this entanglement between the system and the environment is reflected in the system's reduced state, $\hat\rho_\mathrm{s}(t)$, being mixed. For instance, the von Neumann entropy $S_\mathrm{vN}=-\mathrm{Tr}\left[\hat\rho(t)\ln\hat\rho(t)\right]>0$ is used as a measure of the amount of entanglement [33]. From this point forward, we will omit the subscript ``s'' for the reduced density operator of the system.

In the Markovian approximation, as assumed for the LME~(\ref{lind}), this information flow out of the system is forever lost to the environment~\cite{bp}. Furthermore, in deriving equation~(\ref{lind}), we also assume the validity of the Born approximation, which implies that, due to the substantial difference in system sizes, the state of the environment is unaffected by the presence of the small system~\cite{bp}.

As previously mentioned, the loss of information can happen through either dissipation or decoherence. In the former, we often think of particle losses, while in the latter, it typically results from uncontrolled energy shifts. Regardless of the specific mechanism, both processes tend to cause the state of the system to become mixed in most cases. This concept is at the core of fluctuation-dissipation or quantum regression theorems~\cite{bp,mw,carmichael}. However, if one has complete access to the environment, in principle, it is possible to extract all the information about the system. In this scenario, the system's state can be ascribed a pure state. This would be the case of post-selection~\cite{postselect}, briefly touched upon in the final Sec.~\ref{sec:con}.

\section{Liouvillian for non-hermitian SSH model}\label{sec:app2}
It is straightforard to generalize the HN to the Su–Schrieffer–Heeger (SSH) model~\cite{ssh}. The Hamiltonian of the hermitian SSH model is
\begin{equation}\label{sshham}
    \hat H_\mathrm{SSH}=(1-\chi)\sum_{n=1}^N\hat a_n^\dagger\hat b_n+(1+\chi)\sum_{n=1}^N\hat a_{n+1}^\dagger\hat b_n+h.c.
\end{equation}
The parameter $\chi$, which takes values in the interval $[-1,0]$, dictates the relative strengths of consecutive tunneling amplitudes in the SSH model. For $\chi=0$, the model reverts to the conventional tight-binding model. However, for $\chi\neq0$, the lattice exhibits a bipartite structure, with each unit cell containing two distinct types of sites: the $a$-sites and the $b$-sites. In cases with an odd number of sites, the SSH model supports a zero-energy topological edge state, also known as symmetry-protected state, which exhibit exponential localization near the chain's edge. The specific value of $\chi$ determines the size of the energy gap that separates these edge states from the remaining bulk states, as well as the amount of localization of the edge state. These edge states tend to be more isolated from the rest, potentially enhancing their robustness for various applications. Consequently, it is worthwhile to investigate how the findings of the present paper extend to the SSH model. Transitioning to the SSH model from the tight-binding Hamiltonian in Eq.~(\ref{collind}) is straightforward. The results of numerical simulations are illustrated in Fig.~\ref{figa2}, displaying the spectra, scaled positions~(\ref{pos}), and widths 
\begin{equation}\label{wid}
    \Delta\xi=\sqrt{\langle\hat S^2\rangle-\langle\hat S\rangle^2}.
\end{equation}
The plot clearly demonstrates that $\chi$ influences the transitions, shifting them toward smaller values of $\gamma$, and the transition itself becomes smoother. Nonetheless, based on the available numerical data, it appears that the transition remains distinct and is not merely a crossover. A noteworthy observation is that, even when the steady state is not precisely centered in the middle of the lattice, its width remains close to the maximum. Importantly, it should be noted that the model exhibits asymmetry concerning the sign of $\chi$. This asymmetry arises from the drift induced in the lattice by the jump operators, and reversing the sign of $\chi$ results in a distinct model.

\begin{acknowledgements}
The author acknowledges financial support from VR-Vetenskapsr\aa set (The Swedish Research Council), and is thankful for fruitful discussions with Elisabet Edvardsson and Emil Bergholtz. 
\end{acknowledgements}


\begin{thebibliography}{999}
\bibitem{com1} Motivating NH QM as emerging from a more complete description can be done in various ways, and starting from the LME is only one of them. This is, however, the one approach we discuss in this paper.
\bibitem{com2} The degree of localization in the Liouvillian Fock state lattice can be calculated from the inverse partition ratio, e.g. $IPR_j=1/\sum_l|\langle\!\langle l|\varphi_j^R\rangle\!\rangle|^4$, where $|l\rangle\!\rangle$ is a Liouvillian Fock state, and $|\varphi_j^R\rangle\!\rangle$ is the normalized $j$'th eigenvector of the Liouvillian. One finds (not shown in this paper) that all eigenvectors are to some high localized for $\gamma>\gamma_c$, and they are much less localized for $\gamma<\gamma_c$ (even though the limiting value of $IPR_j=N^2$ is never reached even in the delocalized phase).
\bibitem{com3} Photon losses of the laser resonator can be mimicked by a Lindblad jump operator $\hat L=\hat a$, where $\hat a$ is the photon annihilation operator of the light mode. Since a coherent state is an eigenstate of this operator, its action on the state does not alter it; $\hat a|\alpha\rangle=\alpha|\alpha\rangle$. As soon as the state is not a coherent one, the presence of fluctuations will influence it.
\end{thebibliography}
\end{document}